\documentclass[11pt,english,longbibliography]{article}
\usepackage{putex}
\usepackage{latexml}
\iflatexml
  \usepackage{authblk}
\fi

\renewcommand{\Re}{{\rm Re \, }}

\newcommand{\re}{{\rm Re \, }}

\renewcommand{\vec}[1]{\mathbf{#1}}
\newcommand{\CO}{{\cal O}}

\newcommand{\bR}{{\mathbb R}}
\newcommand{\bG}{{\mathbb G}}

\newcommand{\bD}{{\mathbb D}}

\newcommand{\CC}{{\cal C}}

\newcommand{\CF}{{\cal F}}

\newcommand{\dS}{{\text{dS}}}
\newcommand{\CI}{{\cal I}}

\newcommand{\so}{{\mathfrak{so}}}

\newcommand{\CM}{{\cal M}}
\usepackage{amsthm}

\makeatletter
\newcommand*{\rom}[1]{\expandafter\@slowromancap\romannumeral #1@}
\makeatother

\usepackage[latin9]{inputenc}
\usepackage{geometry}
\usepackage{verbatim}
\usepackage{prettyref}
\usepackage[numbers,sort&compress]{natbib}
\usepackage{amsmath}
\usepackage{esint}
\usepackage{amssymb}
\usepackage{braket}
\usepackage{tikz-feynman}
\usetikzlibrary{arrows.meta,calc}
\definecolor{contourblue}{HTML}{16598B}
\definecolor{polered}{HTML}{B23B32}
\definecolor{axisgray}{HTML}{ADB2B8}
\usepackage{tikz-cd}
\tikzset{commutative diagrams/row sep/huge=4cm}
\tikzset{commutative diagrams/column sep/huge=4cm}
\tikzcdset{scale cd/.style={every label/.append style={scale=#1},
    cells={nodes={scale=#1}}}}
\usepackage{lmodern}
\usepackage{tcolorbox}
\usepackage{cases}
\usepackage{bbold}
\usepackage{bm}
\usepackage{ytableau}
\usepackage{youngtab}
\usepackage{mathtools}
\usepackage{graphicx}
\usepackage[nottoc]{tocbibind}
\usepackage{mathrsfs}
\usepackage{caption}
\usepackage{subcaption}
\usepackage{cancel}
\usepackage{float}
\usepackage{babel}
\usepackage{tikz}
\usetikzlibrary{shadings, calc}
\tikzset{
    Witten diagram/.style={
        execute at begin picture={
            \draw[blue, line width=1.5pt] circle[radius=\pgfkeysvalueof{/tikz/Witten/radius}];
            \path node (X){\phantom{X}};
        },
        baseline={(X.base)}
    },
    vertex/.style={circle,fill,inner sep=1.5pt,node contents={}},
    Witten/.cd,
    radius/.initial=3cm
}

\usetikzlibrary{shapes.misc}  
\tikzset{cross/.style={cross out, draw=black, minimum size=2*(#1-\pgflinewidth), inner sep=0pt, outer sep=0pt},
cross/.default={1.5pt}}   
\usetikzlibrary{patterns}

\usetikzlibrary{arrows.meta, decorations.markings}

\tikzset{
  mid arrow/.style={
    postaction={decorate,
      decoration={markings,
        mark=at position 0.5 with {\arrow[#1]{Stealth[length=8pt]}}
      }
    }
  },
  mid arrow/.default={blue!60!black}
}

\tikzset{
  mid arrow rev/.style={
    postaction={decorate,
      decoration={markings,
        mark=at position 0.5 with {\arrowreversed[#1]{Stealth[length=8pt]}}
      }
    }
  },
  mid arrow rev/.default={blue!60!black}
}

\usepackage{lipsum}
\usepackage{framed}
\usepackage{adjustbox}
\usepackage{mathtools}
\usepackage{braket}

\usepackage{slashed}
\usepackage{esvect}
\usepackage{dsfont}

\usepackage{color}
\definecolor{darkgreen}{rgb}{0,0.5,0}
\definecolor{darkblue}{rgb}{0,0,0.6}
\definecolor{purple}{rgb}{0.4,.2,0.7}
\usepackage[colorlinks=true,citecolor=darkgreen,linkcolor=purple,urlcolor=purple]{hyperref}

\numberwithin{equation}{section}
\numberwithin{figure}{section}
\numberwithin{table}{section}

\def\CM{{\cal M}}

\newcommand{\CA}{\mathcal{A}}

\DeclareFontShape{OT1}{cmr}{mx}{n}{<->cmr10}{}
\input{affiliation_icons.tex}

\usepackage{braket}

\begin{document}
\iflatexml
\title{Cosmological correlators from the Euclidean sphere}
\author[1]{Andrei Katsevich}
\author[2]{Zimo Sun}
\affil[1]{Joseph Henry Laboratories and Leinweber Forum for Theoretical Physics,
Princeton University, Princeton, NJ 08544, USA}
\affil[2]{Institute for Advanced Study, Princeton, NJ 08540, USA}
\else
\title{\textbf{Cosmological correlators from\\ 
the Euclidean sphere}}

\authors{Andrei Katsevich\SphereAffil{} and Zimo Sun\dSAffil{}}

\institution{PU}{\AffiliationRow{\SphereAffil{}}{
    Joseph Henry Laboratories and
    Leinweber Forum for Theoretical Physics, Princeton University, Princeton, NJ 08544, USA}}

\institution{IAS}{\AffiliationRow{\dSAffil{}}{Institute for Advanced Study, Princeton, NJ 08540, USA}}

\fi

\abstract{We propose a new framework for computing late-time cosmological correlators in de Sitter  spacetime by analytic continuation from the Euclidean sphere.  In this framework, correlators of arbitrary massive scalar fields take the universal form of generalized Euler integrals. Using this representation, we reproduce the results obtained using the Schwinger-Keldysh formalism for a range of tree-level and loop diagrams, providing nontrivial checks of the equivalence between the Euclidean and the in-in formalisms. We further show that a Landau analysis of the generalized Euler integrals determines the singularities of the corresponding correlators in momentum space. This framework provides a systematic approach to studying the analytic structure of cosmological correlators.}

\date{\today}
\maketitle

\tableofcontents

\newpage
\section{Introduction}
The very early universe is believed to have gone through a period of accelerated expansion, known as inflation. During inflation, the spacetime is approximately de Sitter (dS), and quantum fluctuations are stretched to cosmological scales \cite{Mukhanov:1981xt,Hawking:1982cz,Starobinsky:1982ee,Guth:1982ec,Bardeen:1983qw}. These fluctuations are the origin of the anisotropies in the cosmic microwave background and of the large-scale structure of the universe. Their statistics are described by late-time correlation functions, known as cosmological correlators \cite{Maldacena:2002vr,Planck:2019kim}. The  cosmological correlators encode information about the interactions during inflation, such as the masses and spins of heavy particles \cite{Chen:2009zp,Baumann:2011nk,Noumi:2012vr,Arkani-Hamed:2015bza,Lee:2016vti}. They are also interesting from a theoretical point of view, because they are the dS counterpart of the $S$-matrix in flat spacetime and are constrained by the conformal group  \cite{Witten:2001kn,Strominger:2001pn}. Computing cosmological correlators is nevertheless difficult already at tree level, and many methods have been developed for this purpose. 

\begin{figure}[ht]
\centering
\begin{tikzpicture}[
  font=\small,
  axis/.style={axisgray,thin,-{Stealth[length=1.5mm]}},
  contour/.style={contourblue,line width=.8pt},
  eads/.style={polered,line width=.8pt},
  direction/.style={-{Stealth[length=1.5mm,width=1.2mm]}}
]

\begin{scope}
  \draw[axis] (-3.1,0)--(1.05,0) node[below left,black] {$\operatorname{Re}\eta$};
  \draw[axis] (0,-1.5)--(0,1.6) node[below right,black] {$\operatorname{Im}\eta$};
  \draw[contour] (-2.95,0.41)--(-0.67,0.07)
    arc[start angle=90,end angle=-90,radius=0.07]
    --(-2.95,-0.41);
  \draw[contour,direction] (-1.93,0.258)--(-1.80,0.239);
  \draw[contour,direction] (-1.80,-0.239)--(-1.93,-0.258);
  \fill (-0.60,0) circle[radius=.042];
  \node[below=2pt] at (-0.60,-0.07) {$\eta_0$};
  \node[contourblue] at (-1.85,0.55) {$+$};
  \node[contourblue] at (-1.85,-0.55) {$-$};
  \node at (-1.2,-1.95) {(a) The in-in contour.};
\end{scope}

\begin{scope}[xshift=4.9cm]
  \draw[axis] (-2.7,0)--(1.25,0) node[below left,black] {$\operatorname{Re}\eta$};
  \draw[axis] (0,-1.5)--(0,1.6) node[below right,black] {$\operatorname{Im}\eta$};
  \draw[axisgray,dashed] (-2.5,0.36)--(-0.25,0.04);
  \draw[axisgray,dashed] (-0.25,-0.04)--(-2.5,-0.36);
  \draw[black!55,direction] (172:1.75) arc[start angle=172,end angle=106,radius=1.75];
  \draw[black!55,direction] (188:1.75) arc[start angle=188,end angle=254,radius=1.75];
  \draw[eads] (0,1.35)--(0,0)--(0,-1.35);
  \draw[eads,direction] (0,0.75)--(0,0.62);
  \draw[eads,direction] (0,-0.62)--(0,-0.75);
  \node[polered,right] at (0.05,0.95) {$\eta=iz$};
  \node[polered,right] at (0.05,-0.95) {$\eta=-iz$};
  \node at (-0.85,-1.95) {(b) Rotation to two EAdS.};
\end{scope}

\begin{scope}[xshift=10.2cm]
  \draw[axis] (-2.4,0)--(2.75,0) node[below left,black] {$\operatorname{Re}\eta$};
  \draw[axis] (0,-1.5)--(0,1.6) node[below right,black] {$\operatorname{Im}\eta$};
  \draw[contour] (-1.9,0.285)--(1.75,-0.2625);
  \draw[contour,direction] (-1.07,0.161)--(-0.94,0.141);
  \draw[contour,direction] (0.94,-0.141)--(1.07,-0.161);
  \node[font=\scriptsize] at (-1.2,-0.32) {expanding};
  \node[font=\scriptsize] at (1.2,0.32) {contracting};
  \node at (0,-1.95) {(c) The in-out contour.};
\end{scope}
\end{tikzpicture}
\caption{Time contours for a single vertex in the complex conformal time plane. (a) The time-ordered ($+$) branch runs from $-\infty(1-i\epsilon)$ to some late time $\eta_0\to 0^-$, and the anti-time-ordered ($-$) branch runs back to $-\infty(1+i\epsilon)$. This $i\epsilon$ prescription chooses the Bunch-Davies vacuum. (b) Rotating the two branches to $\eta=\pm iz$, $z>0$, turns each into a copy of EAdS  \cite{Sleight:2020obc,Sleight:2021plv,DiPietro:2021sjt}. (c) The in-out contour of \cite{Donath:2024utn}, which continues the expanding Poincar\'e patch ($\eta<0$) into a contracting one ($\eta>0$) and uses only time-ordered propagators.}
\label{fig:inin}
\end{figure}
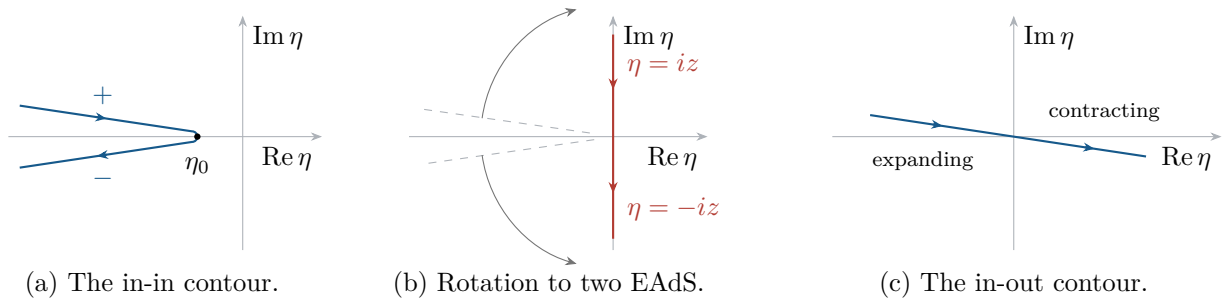

These methods can be roughly divided into two classes. The first class starts from the in-in, or Schwinger--Keldysh formalism \cite{Weinberg:2005vy,Weinberg:2006ac,Adshead:2009cb,Chen:2017ryl}, in which each vertex is integrated along a contour running from the far past (conformal time $\eta=-\infty$) to late time $\eta=0$ and back (Figure~\ref{fig:inin}a). Much of the progress in this approach comes from deforming the contour. Rotating the two branches in opposite directions maps a single in-in diagram to a linear combination of Witten diagrams in Euclidean AdS (EAdS) (Figure~\ref{fig:inin}b) \cite{Sleight:2020obc,Sleight:2021plv,DiPietro:2021sjt}, while gluing a contracting patch to the expanding Poincar\'e patch turns in-in diagrams into in-out diagrams with ordinary Feynman rules (Figure~\ref{fig:inin}c) \cite{Donath:2024utn}. Mellin--Barnes (MB) representations \cite{Sleight:2019mgd,Sleight:2019hfp} and spectral decompositions \cite{Xianyu:2022jwk,Loparco:2023rug,Grafe:2026qsm} further simplify the time integrals, including at loop level \cite{Qin:2023bjk,Qin:2023nhv}.

The second class works directly with boundary data. The cosmological bootstrap determines tree-level correlators from symmetry, locality, unitarity and the structure of their singularities \cite{Arkani-Hamed:2018kmz,Baumann:2019oyu,Goodhew:2020hob,Baumann:2021fxj,Melville:2021lst,Goodhew:2021oqg,Baumann:2022jpr}. For conformally coupled scalars, the wavefunction coefficients, as well as their loop integrands, satisfy differential equations organized by kinematic flow \cite{Arkani-Hamed:2023bsv, Arkani-Hamed:2023kig, Baumann:2024mvm}. For massless spinning fields, their correlators admit a natural formulation on a Grassmannian manifold \cite{Arundine:2026fbr, Arundine:2026myr}.
 
Most of these methods are simplest for conformally coupled or massless scalars, whose mode functions are elementary. For generic masses, the mode functions are Hankel functions, and both the nested time integrals in the in-in formalism and the boundary differential equations become considerably more complicated. Extending the methods above to massive fields is an active direction of research \cite{Gasparotto:2024bku,Herderschee:2025znl,Baumann:2026atn,Arundine:2026myr,Belrhali:2026ygh,Belrhali:2026jqe,Grafe:2026avi,McCune:2026vqb}.
 
\begin{figure}[t]
\centering
\begin{tikzpicture}[
  font=\small,
  axis/.style={axisgray,thin,-{Stealth[length=1.5mm]}},
  contour/.style={contourblue,line width=.8pt},
  eads/.style={polered,line width=.8pt},
  sphere/.style={black,line width=.8pt},
  direction/.style={-{Stealth[length=1.5mm,width=1.2mm]}}
]
\def\h{3.6}  
\def\e{0.09}  

\draw[axis] (-0.5,0)--(6.5,0) node[below left,black] {$\operatorname{Re}\theta$};
\draw[axis] (0,-0.3)--(0,4.15) node[below right,black] {$\operatorname{Im}\theta$};

\draw[black,dashed,thin] (2.5,0)--(2.5,\h);
\node[above] at (2.5,\h) {$\theta=\frac{\pi}{2}+it$};

\draw[eads] (0,0)--(0,\h);
\draw[eads] (5,\h)--(5,0);
\draw[eads,direction] (0,1.75)--(0,1.88);
\draw[eads,direction] (5,1.88)--(5,1.75);
\node[polered,right] at (0.05,2.7) {$\theta=it$};
\node[polered,left] at (4.95,2.7) {$\theta=\pi+it$};

\draw[sphere] (0,-\e)--(5,-\e);
\draw[sphere,direction] (0.85,-\e)--(0.98,-\e);
\draw[sphere,direction] (4.02,-\e)--(4.15,-\e);

\draw[contour,rounded corners=0.18cm] (0,\e)--(2.5-1.4*\e,\e)--(2.5-1.4*\e,\h);
\draw[contour,rounded corners=0.18cm] (2.5+1.4*\e,\h)--(2.5+1.4*\e,\e)--(5,\e);
\draw[contour,direction] (1.55,\e)--(1.68,\e);
\draw[contour,direction] (3.32,\e)--(3.45,\e);
\draw[contour,direction] (2.5-1.4*\e,1.75)--(2.5-1.4*\e,1.88);
\draw[contour,direction] (2.5+1.4*\e,1.88)--(2.5+1.4*\e,1.75);

\fill (0,0) circle[radius=.045];
\fill (5,0) circle[radius=.045];
\node[below left=1pt] at (0,-\e) {$0$};
\node[below=3pt] at (2.5,-\e) {$\pi/2$};
\node[below=3pt] at (5,-\e) {$\pi$};
\end{tikzpicture}
\caption{The complex $\theta$-plane. The black segment $0\le\theta\le\pi$ is the Euclidean sphere and the dashed line $\re\theta=\pi/2$ is (the expanding part of) global dS. The blue lines represent the global in-in contour, with the Euclidean segments preparing the BD vacuum. Along the two red lines the metric is minus that of EAdS$_{d+1}$.}
\label{fig:contour}
\end{figure}
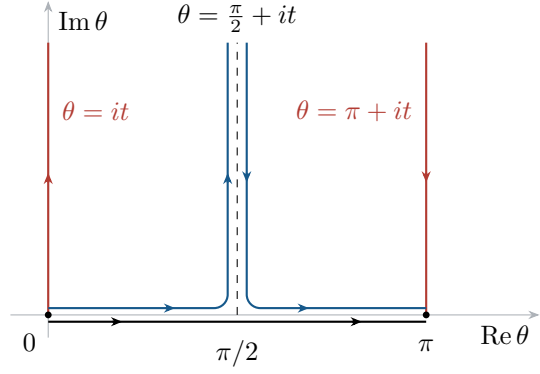

We take a different route, starting from the observation that cosmological correlators are expectation values in the Bunch--Davies (BD) vacuum, which can be prepared by a Euclidean path integral on the Euclidean hemisphere \cite{Bunch:1978yq,Hartle:1983ai}. This suggests a Euclidean calculation on the sphere, which should be equivalent to the in-in formalism. There is an intuitive and perturbative argument for the equivalence based on contour deformations \cite{DiPietro:2021sjt}. Consider the round metric on $S^{d+1}$ \footnote{Throughout the paper, we take the dS radius to be 1, so the Euclidean geometry is a unit sphere.}
\begin{equation}
    ds^2_{S^{d+1}}=d\theta^2+\sin^2\theta\,d\Omega_d^2~,
\end{equation}
where setting $\theta=\frac{\pi}{2}+it$ gives global dS$_{d+1}$. Consider a Feynman diagram, with both interacting vertices and external operator insertions on the sphere. The $\theta$ integral of each vertex then runs along the black contour in Figure~\ref{fig:contour}. As we move the operators to the future boundary of dS, the contour is dragged into the blue one, which consists of both Euclidean and Lorentzian segments. We can further deform it into the red lines $\re\theta=0,\pi$, along which the metric is minus that of EAdS$_{d+1}$. These two lines correspond to the two copies of EAdS obtained by rotating the two branches of the in-in contour (Figure~\ref{fig:inin}b). A more rigorous account of this picture was given by Higuchi, Marolf and Morrison \cite{Higuchi:2010xt}, who argued that, for interacting massive scalars, the in-in correlators agree with the analytically continued sphere correlators diagram by diagram in perturbation theory. \footnote{This agreement has been questioned due to infrared divergences, even for massive scalars \cite{Polyakov:2007mm, Polyakov:2009nq, Krotov:2010ma, Polyakov:2012uc}. As pointed out in \cite{Marolf:2011sh}, the discrepancy is due to the choice of initial conditions.}
 
Despite the equivalence between the two formalisms, the sphere has not been used as a practical tool for computing late-time correlators. The main obstacle is that Feynman integrals on the sphere are themselves hard to evaluate. Expanding the propagators in spherical harmonics leads to complicated infinite sums, which are difficult to resum and do not admit a simple analytic continuation to dS. Some earlier attempts include deriving MB representations of sphere propagators in position space \cite{Marolf:2010nz,Hollands:2010pr}. But these methods have not been applied to study properties of  late-time correlators.

To overcome this difficulty, we use the embedding space framework proposed in \cite{Bandaru:2024qvv}, which transforms Feynman diagrams on the sphere into Schwinger parameter integrals. In this framework, every diagram becomes a generalized Euler integral, with two Schwinger parameters associated with each propagator. The masses only enter through the exponents of the Schwinger parameters, so fields of arbitrary mass are treated on an equal footing. We review the embedding space Feynman rules on the sphere in Section~\ref{Frules}. The analytic continuation to the late-time boundary of dS can be performed directly on the integrand. We illustrate this procedure in Section~\ref{sec:S-in} for contact, exchange and $n$-loop candy diagrams, and show that taking the late-time limit reduces the number of Schwinger parameters on each external line to one. Using the resulting Euler integrals in position space, we explicitly verify the equivalence between the sphere and in-in calculations, for the candy diagrams at arbitrary loop order.

Generalized Euler integrals for scalars of arbitrary mass have also been obtained in \cite{Herderschee:2025znl} by an entirely different approach, which works directly in the in-in formalism, so that each branch assignment of the in-in contour gives a separate integral. In our framework, the sum over all branch assignments is instead captured by a single integral, so that the full correlator can be studied directly. In Section~\ref{sec:exch}, we Fourier transform the exchange diagram to momentum space and use its Euler integral representation to study the collapsed limit, and recover the Mellin representation of \cite{Sleight:2019mgd}. In Section~\ref{sec:Landau}, by applying a Landau analysis to the momentum-space integrals, we recover all the singularities of contact, exchange and candy diagrams as solutions of the Landau equations.

\subsection{Notations and conventions}
Let $X^I, I=0,1, \cdots, d+1$ be the Cartesian coordinates of the embedding space $\mathbb{R}^\bD$, where $\bD=d+2$. The unit sphere $S^{d+1}$ in the embedding space  corresponds to unit vectors $\hat X^I = X^I/|X|$. Wick rotating the $\hat X^0$ component of $S^{d+1}$ to the imaginary axis yields the embedding space representation of dS$_{d+1}$ with unit radius: $-Y_0^2+ Y_1^2+\cdots Y_{d+1}^2 = 1$. We shall consider the Poincar\'e coordinates  of dS$_{d+1}$ 
\begin{equation}\label{flatslicing}
    Y^I = \left(-\frac{1+\vec y^2-\eta^2}{2\eta}, -\frac{\vec y}{\eta}, \frac{1-\vec y^2+\eta^2}{2\eta}\right), \quad  \vec y \in \bR^d~,
\end{equation}
where $\eta<0$, with $\eta=0$ corresponding to the future boundary. The dS metric in this  coordinate system reads $ds^2 = \frac{-d\eta^2+d\vec y^2}{\eta^2}$.

Given a free massive scalar field $\phi$ in dS$_{d+1}$, we parameterize its mass by a scaling dimension $\Delta\in (0,d/2) \cup \frac{d}{2}+i\bR_{\ge 0}$ via the standard relation $m^2=\Delta(d-\Delta)$. Following the terminology of CFT literature, we define the ``shadow dimension'' $ \bar\Delta \equiv d-\Delta$. The leading boundary behavior of $\phi$ is given by 
\begin{equation}\label{latetime}
    \phi(\eta, \vec y)\stackrel{\eta\to 0}{\approx} (-\eta)^\Delta \CO_{\Delta} (\vec y) +(-\eta)^{\bar\Delta} \CO_{\bar\Delta}(\vec y)\,,
\end{equation}
where $\CO_{\Delta}$ and $\CO_{\bar\Delta}$ transform as scalar primary operators of dimensions $\Delta$ and $\bar\Delta$, respectively, under $\so(1, d+1)$. A scalar field belongs to the principal series when $m\ge \frac{d}{2}$ or equivalently when $\Delta=\frac{d}{2}+i\nu, \nu\ge 0$. It belongs to the complementary series when $0<m< \frac{d}{2}$ or equivalently when $0<\Delta<\frac{d}{2} $. In our convention, we always have $\Delta<\bar\Delta$ in the complementary series.

We will use the following shorthand notations
\begin{equation}\label{intshort}
   \int_{\hat X}\equiv \int_{S^{d+1}} d^{d+1}\hat X, \quad  \int_X \equiv \int_{\bR^\bD} d^\bD X, \quad \int_\lambda = \int_0^\infty \frac{d\lambda}{\lambda}\,.
\end{equation}

\section{Feynman integrals on the sphere}\label{Frules}
As a spherical counterpart of the Fourier space in flat space, the harmonic basis provides a natural way to compute Feynman diagrams on the sphere. However, it is technically challenging to obtain good analytical control of integrals of the spherical harmonics and the resulting complicated infinite sums. Moreover, the harmonic basis does not admit a simple analytic continuation to dS. To overcome these technical difficulties, we will use the embedding space framework proposed in \cite{Bandaru:2024qvv} to transform Feynman diagrams on the sphere into Schwinger parameter integrals. In this section, we review this embedding space approach, which consists of two parts: lifting the propagators to  embedding space, and lifting the vertex integrals to  embedding space.

First, for a scalar field $\phi$ on $S^{d+1}$ of mass $m^2=\Delta\bar\Delta$, the corresponding two-point function is 
\begin{equation}\label{GF}
    G_\Delta (\hat X_1, \hat X_2) =\frac{\Gamma(\Delta)\Gamma(\bar\Delta)}{(4\pi)^{\frac{d+1}{2}}} \mathbf F\left(\Delta,\bar\Delta;\frac{d+1}{2};\frac{1+\hat X_1\cdot \hat X_2}{2}\right)\,,
\end{equation}
where $\mathbf F(a,b;c;z)={}_2F_1(a,b;c;z)/\Gamma(c)$ is the regularized hypergeometric function. It was noticed in \cite{Bandaru:2024qvv} that  $G_\Delta (\hat X_1, \hat X_2) $ admits a simple integral representation\footnote{The same integral representation can be extracted from Appendix B.1 of \cite{Penedones:2010ue}, there for EAdS harmonic functions. Up to normalization, those are given by the same hypergeometric function as \eqref{GF}, with $\hat X$ read as an EAdS embedding coordinate, satisfying $\hat X^2=-1$.}
\begin{equation}\label{Gi}
    G_\Delta (\hat X_1, \hat X_2) = \frac{1}{2\pi^{\bD/2}}\int_{\lambda, \mu }\lambda^{\bar\Delta}\mu^{\Delta} e^{-(\lambda \hat X_1-\mu\hat X_2)^2}\,,
\end{equation} 
where the integration measures for the Schwinger parameters $\lambda, \mu$ are defined in \eqref{intshort}. The integral representation allows for a natural extension to embedding space. Replacing $\hat{X} \in S^{d+1}$ with $X \in \bR^{\bD}$ in $G_\Delta$, we obtain the extended two-point function $\bG_\Delta(X_1, X_2)$, which satisfies the asymmetric  homogeneity condition
\begin{equation}\label{homo}
    \bG_\Delta(s X_1, t X_2)= s^{-\bar\Delta} t^{-\Delta} \bG_\Delta(X_1, X_2),\quad s,t>0\,.
\end{equation}
Since the two points have different properties under rescaling, we represent the propagator diagrammatically by an arrow
\begin{equation}
    \bG_\Delta( X_1, X_2) = \vcenter{\hbox{
\begin{tikzpicture}
  \draw[->, blue] (0,0) -- (1,0);
  \draw [blue] (1,0) -- (2,0);
  \node at (1,0.4) {\(\Delta\)};
  \filldraw (0,0) circle (1.5pt) node[left] {\( X_1\)};
  \filldraw (2,0) circle (1.5pt) node[right] {\( X_2\)};
\end{tikzpicture}
}}
\end{equation}
In this convention, the tail of the arrow is associated with homogeneity degree $-\bar\Delta$ while the head is associated with homogeneity degree $-\Delta$.

The advantage of $\bG_\Delta(X_1, X_2)$ is that its coordinate dependence is Gaussian. So, if we can replace $\int_{\hat{X}}$ by $\int_X$, the vertex integrals become straightforward. This is achieved by a trick introduced in \cite{Bandaru:2024qvv}. We illustrate this with the $n$-point contact diagram as an example. The contact diagram is shown in Figure \ref{fig:contact}, with $\hat X_0$ being the integration vertex and the remaining $\hat X_i$ being external points. The sphere integral reads
\begin{equation}
    I_\text{cont}(\hat X_i) = \int_{\hat X_0} \prod_{i=1}^n G_{\Delta_i}(\hat X_i, \hat X_0)\,.
\end{equation}
We lift $\hat X_0$ to embedding space and we choose all arrows to point towards $X_0$. The key observation is that each extended propagator $\bG_{\Delta_i}(\hat X_i, X_0)$ has a definite homogeneity degree in $X_0$ given by \eqref{homo}. So the full integrand scales as $t^{-\Delta_\text{tot}}$ under $X_0 \to tX_0$, where $\Delta_{\rm tot}=\sum_{i=1}^n\Delta_i$. This allows us to trade the integral over $S^{d+1}$ for an integral over $\mathbb{R}^{\bD}$, at the cost of introducing an $\bR_+$ gauge redundancy:
\begin{equation}
    I_\text{cont}(\hat X_i) = \frac{1}{\mathrm{Vol}(\bR_+)} \int_{X_0} \; |X_0|^{\Delta_\text{tot} - \bD} \prod_{i=1}^n G_{\Delta_i}(\hat X_i, X_0)\,.
\end{equation}
The factor $|X_0|^{\Delta_\text{tot} - \bD}$ compensates for the homogeneity of the integrand, making the full expression invariant under the ``gauge'' transformation $X_0 \to tX_0$. The gauge group volume  is defined by ${\rm Vol}(\bR_+) = \int_0^\infty dt/t$. We exponentiate $|X_0|^{\Delta_\text{tot} - \bD}$ using a Schwinger parameter $\tau$ \footnote{Here we assume $\bD>\Re(\Delta_{\rm tot})$. The final result can be analytically continued to more general $\Delta_i$.}, which allows us to rewrite $I_\text{cont}(\hat X_i) $ as follows
\begin{equation}
    I_\text{cont}(\hat X_i) = \frac{1}{\mathrm{Vol}(\bR_+)} \frac{(2\pi^{\bD/2})^{-n}}{\Gamma\!\left(\frac{\bD - \Delta_\text{tot}}{2}\right)}\int_{\lambda,\mu,\tau} \tau^{\frac{\bD-\Delta_\text{tot}}{2}} \prod_i\left(\lambda_i^{\bar\Delta_i}\mu_i^{\Delta_i}\right)\int_{X_0} e^{-\tau X_0^2-\sum_i (\lambda_i \hat X_i-\mu_i X_0)^2}\,.
\end{equation}
This integral has the gauge symmetry $(\mu_i, \tau, X_0) \rightarrow (\mu_i/t, \tau/t^2, t X_0) $. Following the standard Faddeev-Popov (FP) procedure, we fix the gauge by choosing $\tau=1$, and insert the corresponding FP determinant $\Delta_{\rm FP} = 2$. We can also integrate out $X_0$, which is purely Gaussian
\begin{equation}\label{Icont}
   I_\text{cont}(\hat X_i) =  \frac{(2\pi^{\bD/2})^{1-n}}{\Gamma\!\left(\frac{\bD - \Delta_\text{tot}}{2}\right)}\int_{\lambda,\mu}\frac{\prod_i\lambda_i^{\bar\Delta_i}\mu_i^{\Delta_i}}{(1+\mu^2)^{\bD/2}}\exp\left(- \frac{\lambda^2\!+\!\lambda^2\mu^2\!-\!(\lambda\cdot\mu)^2+2\sum_{i<j}\lambda_i\lambda_j\mu_i\mu_j(1\!-\!\hat X_i\cdot \hat X_j))}{1 + \sum_i \mu_i^2}\right)\,.
\end{equation}
Here, we adopt the shorthand notations $\lambda^2=\sum_i\lambda_i^2$, $\mu^2=\sum_i\mu_i^2$, and  $\lambda\cdot\mu=\sum_i\lambda_i\mu_i$.

This formalism admits a straightforward generalization to derivative interactions. We work out an explicit example in Appendix \ref{derivative}.

\section{The equivalence between the sphere and in-in formalisms}\label{sec:S-in}
Building on the integral representations reviewed in the previous section, we show how to analytically continue correlation functions on the sphere to dS and take the late-time limit. We explicitly verify that, for a class of tree-level and loop diagrams, the late-time correlators obtained in this way precisely match the corresponding in-in correlators in dS.

\subsection{Contact diagrams}\label{subseccontact}

\begin{figure}
    \centering
\begin{tikzpicture}[scale=1]

  \shade[ball color=blue!10!white, opacity=0.5] (-4.5, 0) circle (2cm);
  \draw[thick] (-4.5, 0) circle (2cm);

  \draw[dashed, gray, thick] (-4.5, 0) ellipse (2cm and 0.6cm);

  \coordinate (V1) at (-4.5, -0.6);

  \coordinate (S1) at (-5.8,  0.2);
  \coordinate (S2) at (-5.2,  1.2);
  \coordinate (S3) at (-4.4,  1.55);
  \coordinate (Sn) at (-3.2,  0.2);

  \foreach \P in {S1, S2, S3, Sn}{
    \draw[thick, blue!60!black, mid arrow rev] (V1) -- (\P);
  }

  \node[left]       at (S1) {$\hat X_1$};
  \node[left] at (S2) {$\hat X_2$};
  \node[right]      at (S3) {$\hat X_3$};
  \node[right]      at (Sn) {$\hat X_n$};

  \foreach \P in {S1, S2, S3, Sn}{
    \filldraw[red] (\P) circle (2pt);
  }

  \filldraw[black] (V1) circle (2.5pt);
  \node[below, font=\small] at (V1) {$\hat{X}_0$};

  \node[font=\normalsize] at (-3.75, 0.85) {$\cdots$};

  \node[font=\small, gray] at (-4.5, -2.6) {$S^{d+1}$};

  \draw[thick, ->] (0.3, -2) -- (0.3, 2);

  \draw[thick, gray] (0.0, 1.6) -- (7.0, 1.6);
  \node[left, font=\small] at (0.0, 1.6) {$\eta\!=\!0$};

  \coordinate (P1) at (1.0, 1.6);
  \coordinate (P2) at (2.4, 1.6);
  \coordinate (P3) at (3.8, 1.6);
  \coordinate (Pn) at (6.0, 1.6);

  \coordinate (V2) at (3.3, -1.8);

  \foreach \P in {P1, P2, P3, Pn}{
    \draw[thick] (V2) -- (\P);
  }

  \foreach \P in {P1, P2, P3, Pn}{
    \filldraw[red] (\P) circle (2pt);
  }

  \filldraw[black] (V2) circle (2.5pt);

  \node[above, font=\small] at (P1) {$\vec y_1$};
  \node[above, font=\small] at (P2) {$\vec y_2$};
  \node[above, font=\small] at (P3) {$\vec y_3$};
  \node[above, font=\small] at (Pn) {$\vec y_n$};

  \node[font=\normalsize] at (4.4, 0.9) {$\cdots$};

  \node[font=\small, gray] at (3.5, -2.6) {$\mathrm{dS}_{d+1}$};

\end{tikzpicture}
    \caption{An $n$-point contact diagram on the sphere and in dS. A black dot denotes an internal vertex and a red dot represents an external operator insertion.}
    \label{fig:contact}
\end{figure}
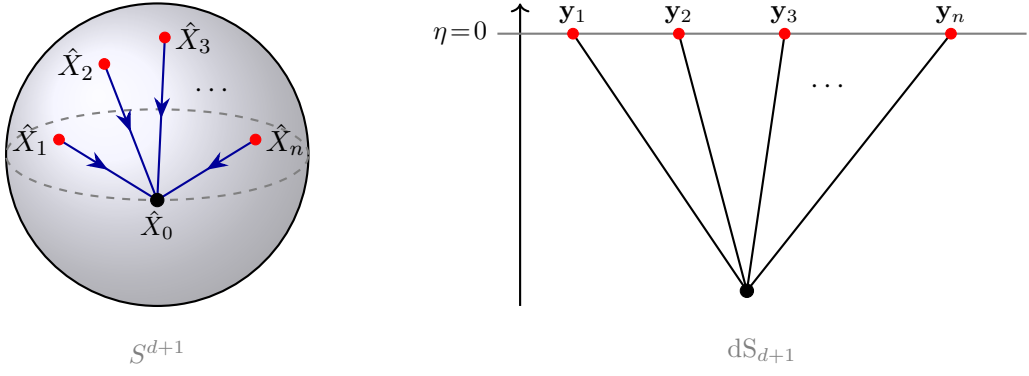

For the contact diagram on the sphere, we have derived the integral representation~\eqref{Icont}. The resulting expression is a function of $\hat X_i\cdot \hat X_j$, and the defining integral converges due to the bound $\hat X_i \cdot \hat X_j\le 1$ on $S^{d+1}$. Convergence is preserved as we analytically continue the external points from $S^{d+1}$ to $\mathrm{dS}_{d+1}$, provided they remain spacelike separated. Denoting by $Y_i$ the image of $\hat X_i$ under this analytic continuation, in the flat slicing~\eqref{flatslicing} we have
\begin{equation}\label{Wick}
    2(1-\hat X_i\cdot \hat X_j) \;\to\; 2(1-Y_i\cdot Y_j)=\frac{\vec y_{ij}^2-\eta_{ij}^2}{\eta_i\eta_j}\,,
\end{equation}
where $\vec y_{ij}=\vec y_i-\vec y_j$, $\eta_{ij}=\eta_i-\eta_j$. Together with the Wick rotation \eqref{Wick}, equation \eqref{Icont} therefore defines an $n$-point correlation function $I_{\rm cont}(Y_i)$ on $\mathrm{dS}_{d+1}$, subject to the constraints $|\vec y_{ij}|\ge |\eta_{ij}|$. In particular, these constraints are automatically satisfied in the late-time limit.

We now explain how to extract the late-time cosmological correlators following the Wick rotation to dS. The late-time limit of a free field is given by \eqref{latetime}. Focusing on the complementary series with the convention $\Delta<\bar\Delta$, the leading late-time behavior of a bulk correlation function is captured by $\prod_{i}(-\eta_i)^{\Delta_i}\langle\prod_i \mathcal{O}_{\Delta_i}(\vec y_i)\rangle$. To match the powers of $\eta_i$, we perform the rescaling $\mu_i\to (-\eta_i)\mu_i$ in $I_{\rm cont}(Y_i)$. The factor $\int_\mu \prod_i\mu_i^{\Delta_i}$ then produces $\prod_{i}(-\eta_i)^{\Delta_i}$ upon rescaling, and for the remaining part of the integrand we may safely take the limit $\eta_i\to 0$, yielding
\begin{equation}\label{OS1}
    \langle \mathcal{O}_{\Delta_1}(\vec y_1)\cdots \mathcal{O}_{\Delta_n}(\vec y_n)
    \rangle_{\rm cont}^{\SphereAffil{}} 
    = \frac{(2\pi^{\bD/2})^{1-n}}
    {\Gamma\!\left(\frac{\bD - \Delta_\text{tot}}{2}\right)}
    \int_{\lambda,\mu} \prod_i \lambda_i^{\bar\Delta_i}\mu_i^{\Delta_i} 
    \exp\!\left(-\lambda^2-\sum_{i<j}
    \lambda_i\lambda_j\mu_i\mu_j \vec y_{ij}^2\right)~.
\end{equation}
Throughout the paper, we use the notation $\langle\cdots\rangle\SphereAffil{}$ for correlation functions computed using the sphere. Decoupling $\lambda$ and $\mu$ through the substitution $\mu_i\to \mu_i/\lambda_i$, the $\lambda$ integrals become straightforward and yield
\begin{equation}\label{Icont2}
    \langle \mathcal{O}_{\Delta_1}(\vec y_1)\cdots \mathcal{O}_{\Delta_n}(\vec y_n)\rangle_{\rm cont}^{\SphereAffil{}}=\frac{\prod_i \Gamma\left(\frac{d}{2}-\Delta_i\right)}{2^n \left(2\pi^{\frac{\bD}{2}}\right)^{n-1}\Gamma\left(\frac{\bD-\Delta_{\rm tot}}{2}\right)}D_{\Delta_1,\cdots, \Delta_n}(\vec y_i)\,,
\end{equation}
where $D_{\Delta_1,\cdots, \Delta_n}(\vec y_i)=\int_\mu \prod_i\mu_i^{\Delta_i} e^{-\sum_{i<j}\mu_i \mu_j \vec y_{ij}^2}$ is the standard $D$-function in EAdS \cite{DHoker:1999kzh}. The result \eqref{Icont2} can then be analytically continued to all $\Delta_i$, including both principal and complementary series.

The corresponding contact Witten diagram in 
$\mathrm{EAdS}_{d+1}$ reads
\begin{equation}\label{Wcont}
    W^{\rm AdS}_{\rm cont} (\vec y_i ) = \frac{\Gamma\!\left(\frac{\Delta_{\rm tot}-d}{2}\right)D_{\Delta_1,\cdots, \Delta_n}(\vec y_i)}{2(2\pi^{\frac{d}{2}})^{n-1}\prod_i \Gamma\!\left(\Delta_i-\frac{d}{2}+1\right)}\,,
\end{equation}
where we follow the normalization of \cite{Sleight:2021plv} for the AdS boundary-to-bulk propagators. Taking the ratio of \eqref{Icont2} and \eqref{Wcont}, we arrive at
\begin{equation}\label{contactpos}
    \langle \mathcal{O}_{\Delta_1}(\vec y_1)\cdots \mathcal{O}_{\Delta_n}(\vec y_n)\rangle_{\rm cont}^{\SphereAffil{}} = 2\sin\!\left(\frac{\Delta_{\rm tot}\!-\!d}{2}\pi\right)\left(\prod_i c_{\Delta_i}\right)W^{\rm AdS}_{\rm cont} (\vec y_i )\,,\quad c_{\Delta}\equiv\frac{1}{2\sin(\pi(\frac{d}{2}-\Delta))}\,.
\end{equation}
In \cite{Sleight:2021plv}, the contact diagram (cf.\ the right panel of Figure \ref{fig:contact}) was computed by rotating the in-in contour to EAdS, and their result is in precise agreement with~\eqref{contactpos}.

\subsection{Exchange diagrams}

\begin{figure}
    \centering
    \begin{tikzpicture}[scale=1]

  \shade[ball color=blue!10!white, opacity=0.5] (-4.5, 0) circle (2cm);
  \draw[thick] (-4.5, 0) circle (2cm);

  \draw[dashed, gray, thick] (-4.5, 0) ellipse (2cm and 0.6cm);

  \coordinate (V1) at (-4.5,  0.75);
  \coordinate (V2) at (-4.5, -0.75);

  \draw[thick, blue!60!black,mid arrow] (V1) -- (V2);

  \coordinate (S1) at (-5.5,  1.2);   
  \coordinate (S2) at (-3.5,  1.2);   
  \coordinate (S3) at (-5.5, -1.2);   
  \coordinate (S4) at (-3.5, -1.2);   

  \draw[thick, blue!60!black, mid arrow rev] (V1) -- (S1);
  \draw[thick, blue!60!black,mid arrow rev] (V1) -- (S2);
  \draw[thick, blue!60!black, mid arrow rev] (V2) -- (S3);
  \draw[thick, blue!60!black, mid arrow rev] (V2) -- (S4);

  \node[above right]  at (S1) {$\hat X_1$};
  \node[above left] at (S2) {$\hat X_2$};
  \node[below right]  at (S3) {$\hat X_3$};
  \node[below left] at (S4) {$\hat X_4$};

  \foreach \P in {S1, S2, S3, S4}{
    \filldraw[red] (\P) circle (2pt);
  }

  \filldraw[black] (V1) circle (2.5pt);
  \filldraw[black] (V2) circle (2.5pt);

  \node[font=\small, gray] at (-4.5, -2.6) {$S^{d+1}$};

  \draw[thick, ->] (0.3, -2) -- (0.3, 2);

  \draw[thick, gray] (0.0, 1.6) -- (7.2, 1.6);
  \node[left, font=\small] at (0.0, 1.6) {$\eta\!=\!0$};

  \coordinate (P1) at (1.0, 1.6);
  \coordinate (P2) at (2.5, 1.6);
  \coordinate (P3) at (4.2, 1.6);
  \coordinate (P4) at (5.8, 1.6);

  \coordinate (W1) at (1.75, -1.6);
  \coordinate (W2) at (4.75, -1.6);

  \draw[thick] (W1) -- (W2);

  \draw[thick] (W1) -- (P1);
  \draw[thick] (W1) -- (P2);
  \draw[thick] (W2) -- (P3);
  \draw[thick] (W2) -- (P4);

  \foreach \P in {P1, P2, P3, P4}{
    \filldraw[red] (\P) circle (2pt);
  }

  \filldraw[black] (W1) circle (2.5pt);
  \filldraw[black] (W2) circle (2.5pt);

  \node[above, font=\small] at (P1) {$\vec y_1$};
  \node[above, font=\small] at (P2) {$\vec y_2$};
  \node[above, font=\small] at (P3) {$\vec y_3$};
  \node[above, font=\small] at (P4) {$\vec y_4$};

  \node[font=\small, gray] at (3.5, -2.6) {$\mathrm{dS}_{d+1}$};

\end{tikzpicture}
    \caption{An $s$-channel four-point exchange diagram on the sphere and in dS.}
    \label{fig:exchange}
\end{figure}
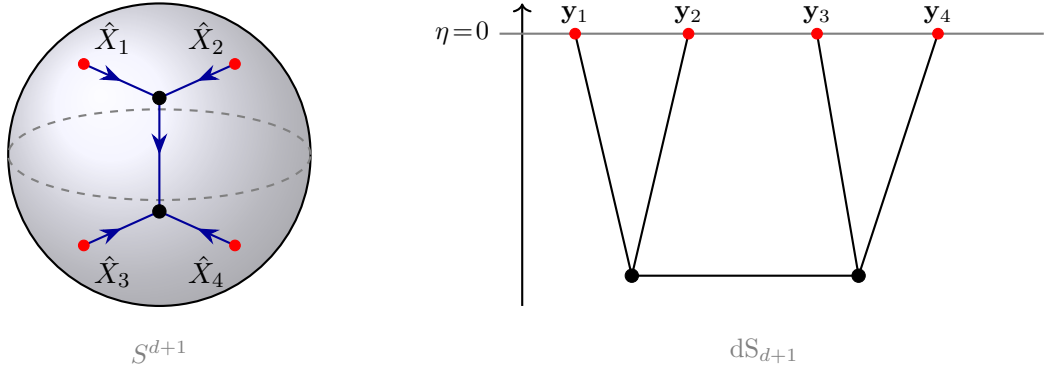

Next, we consider the 4-point exchange diagram on the sphere and in dS, as shown in Figure \ref{fig:exchange}. Each external point $\hat X_i$ on the sphere  is associated with a free propagator of dimension $\Delta_i$, and the internal line has dimension $\Delta$. Integrating out the two internal points following the Feynman rules reviewed in Section \ref{Frules}, the exchange diagram on the sphere can be expressed as\footnote{This specific integral representation relies on the choice of arrows in Figure \ref{fig:exchange}. But as emphasized before, the final result is independent of such a choice.}
\begin{equation}
    \CI_{\rm ex} (\hat X_i)= \frac{\int_{\lambda,\mu,\rho,\sigma}\prod_i(\lambda_i^{\bar\Delta_i}\mu_i^{\Delta_i})\rho^{\bar\Delta}\sigma^\Delta   e^{ -(\lambda^2-W^T\CA^{-1} W)}}{(2\pi^{\bD/2})^{3} \Gamma(\frac{\bD-\Delta_L}{2})\Gamma(\frac{\bD-\Delta_R}{2})\det(\CA)^{\bD/2}}\,,
\end{equation}
where $\Delta_L = \Delta_1+\Delta_2+\bar\Delta$, $\Delta_R = \Delta_3+\Delta_4+\Delta$, and 
\begin{equation}\label{VWA}
    W =
    \begin{pmatrix}
    \lambda_1\mu_1\hat X_1+\lambda_2\mu_2\hat X_2\\ 
    \lambda_3\mu_3\hat X_3+\lambda_4\mu_4\hat X_4
    \end{pmatrix},  \quad \CA =
    \begin{pmatrix} 
    1+\rho^2+\mu_1^2+\mu_2^2 & -\rho\sigma\\
    -\rho\sigma & 1+\sigma^2+\mu_3^2+\mu_4^2 \end{pmatrix}\,.
\end{equation}
It is straightforward to check that the exponent $\lambda^2-W^T\CA^{-1} W$ is manifestly positive definite provided $\hat X_i\cdot \hat X_j\le 1$. This feature is expected to hold for arbitrary massive diagrams on the sphere in dimensional regularization since there is no divergence on the sphere\footnote{See \cite{Bandaru:2024qvv} for an explicit proof of this statement in the case of vacuum diagrams.}. The positivity allows us to analytically continue points on the sphere to spacelike separated configurations in dS. Taking the boundary limit by rescaling $\mu_i\rightarrow(-\eta_i)\mu_i/\lambda_i$ as in the contact diagram case, we obtain
\begin{align}
    &\langle \CO_{\Delta_1}(\vec y_1)\cdots \CO_{\Delta_4}(\vec y_4) \rangle_{\rm ex}^{\SphereAffil{}}= \frac{1}{(2\pi^{\bD/2})^{3} \Gamma(\frac{\bD-\Delta_L}{2})\Gamma(\frac{\bD-\Delta_R}{2})} \int_{\lambda,\mu,\rho,\sigma}\frac{\prod_i(\lambda_i^{\bar\Delta_i-\Delta_i}\mu_i^{\Delta_i})\rho^{\bar\Delta}\sigma^\Delta}{(1+\rho^2+\sigma^2)^{\bD/2}}e^{-\lambda^2-\frac{\Phi_\text{ex}(\vec y_i)}{1+\rho^2+\sigma^2}}\,,\nonumber\\
    &\Phi_{\rm ex}(\vec y_i)\equiv (1\!+\!\sigma^2)\mu_1\mu_2 \vec y_{12}^2\!+\!(1\!+\!\rho^2)\mu_3\mu_4\vec y_{34}^2+\!\rho\sigma(\mu_1\mu_3 \vec y_{13}^2\!+\!\mu_1\mu_4 \vec y_{14}^2\!+\!\mu_2\mu_3 \vec y_{23}^2\!+\!\mu_2\mu_4\vec  y_{24}^2)\,.
\end{align}
Then we integrate out all $\lambda_i$ and make the substitution $\mu_i\to \mu_i\sqrt{1+\rho^2+\sigma^2}$ to remove the overall factor $1+\rho^2+\sigma^2$ in the exponent:
\begin{equation}\label{ss}
    \langle \CO_{\Delta_1}(\vec y_1)\cdots \CO_{\Delta_4}(\vec y_4) \rangle_{\rm ex}^{\SphereAffil{}}=\frac{\CC_{\rm ex}}{ 2^4(2\pi^{\bD/2})^{3}}\int_{\rho, \sigma}\frac{\rho^{\bar\Delta}\sigma^\Delta }{(1\!+\!\rho^2\!+\!\sigma^2)^{\frac{\bD-\Delta_{\rm tot}}{2}}}\int_\mu \prod_i\mu_i^{\Delta_i} e^{-\Phi_{\rm ex} }\,,
\end{equation}
where 
\begin{equation}
    \CC_{\rm ex}= \frac{\prod_i \Gamma(\frac{d}{2}-\Delta_i)}{\Gamma\left(\frac{\bD-\Delta_L}{2}\right)\Gamma\left(\frac{\bD-\Delta_R}{2}\right)}\,.
\end{equation}
The same $\mu_i$ integrals also appear in the corresponding exchange Witten diagram in EAdS \cite{Penedones:2010ue}, but the $\rho$ and $\sigma$ integrals are new.

Our goal is to match the integral representation~\eqref{ss} with the in-in correlator shown in the right panel of Figure~\ref{fig:exchange}. In the in-in formalism, each internal vertex can be placed on either branch of the contour, leading to four contributions in total. Among these, only two are inequivalent. After rotating the in-in contour to EAdS, the correlator can be expressed as a linear combination of two $s$-channel exchange Witten diagrams, with either $\Delta$ or $\bar\Delta$ boundary conditions imposed on the internal line \cite{Sleight:2021plv}. In Appendix \ref{mellin}, we show that the Mellin amplitude of $\langle \CO_{\Delta_1}(\vec y_1)\cdots \CO_{\Delta_4}(\vec y_4) \rangle_{\rm ex}^{\SphereAffil{}}$, defined via \eqref{Mder}, indeed splits into two terms:
\begin{equation}\label{Mrelation}
    \CM^{\rm dS}_{\rm ex} = 4\left(\prod_{i}c_{\Delta_i}\right)\left(c_{\Delta}\sin\left(\frac{\Delta_1\!+\!\Delta_2\!-\!\bar\Delta}{2}\pi\right)\sin\left(\frac{\Delta_3\!+\!\Delta_4\!-\!\bar\Delta}{2}\pi\right) \CM^{\rm AdS}_{\rm ex, \Delta}+(\Delta\leftrightarrow\bar\Delta)\right)\,.
\end{equation}
Here $\CM^{\rm AdS}_{\rm ex,\Delta}$ denotes the EAdS Mellin amplitude with $\Delta$ boundary condition imposed on the exchanged field. With our choice of  normalization, the relation~\eqref{Mrelation} continues to hold if the dS Mellin amplitude $\CM^{\rm dS}_{\rm ex}$ is replaced by the position-space correlator $\langle \CO_{\Delta_1}(\vec y_1)\cdots \CO_{\Delta_4}(\vec y_4)\rangle_{\rm ex}^{\SphereAffil{}}$, and the EAdS Mellin amplitudes are replaced by the corresponding Witten diagrams. We therefore find precisely the same linear combination of EAdS exchange Witten diagrams as in the in-in computation of~\cite{Sleight:2021plv}, including all numerical prefactors. This establishes the agreement between the sphere computation and the in-in result for the exchange diagram.

\subsection{Candy diagrams}\label{sec:candy}

\begin{figure}
    \centering
\begin{tikzpicture}[scale=1]

  \shade[ball color=blue!10!white, opacity=0.5] (-4.5, 0) circle (2cm);
  \draw[thick] (-4.5, 0) circle (2cm);
  \draw[dashed, gray, thick] (-4.5, 0) ellipse (2cm and 0.6cm);

  \coordinate (V1) at (-4.5,  0.75);
  \coordinate (V2) at (-4.5, -0.75);

  \draw[thick, blue!60!black] (V1) to[bend right=50] (V2);  
   \draw[thick, blue!60!black] (V1) to[bend right=15] (V2); 
  \node[font=\small, blue!60!black] at (-4.37, 0.0) {$\cdots$};
  \draw[thick, blue!60!black] (V1) to[bend left=50]  (V2); 

  \coordinate (S1) at (-5.5,  1.2);
  \coordinate (S2) at (-3.5,  1.2);
  \coordinate (S3) at (-5.5, -1.2);
  \coordinate (S4) at (-3.5, -1.2);

  \draw[thick, blue!60!black] (V1) -- (S1);
  \draw[thick, blue!60!black] (V1) -- (S2);
  \draw[thick, blue!60!black] (V2) -- (S3);
  \draw[thick, blue!60!black] (V2) -- (S4);

  \node[above right]  at (S1) {$\hat X_1$};
  \node[above left] at (S2) {$\hat X_2$};
  \node[below right]  at (S3) {$\hat X_3$};
  \node[below left] at (S4) {$\hat X_4$};

  \foreach \P in {S1, S2, S3, S4}{
    \filldraw[red] (\P) circle (2pt);
  }

  \filldraw[black] (V1) circle (2.5pt);
  \filldraw[black] (V2) circle (2.5pt);

  \node[font=\small, gray] at (-4.5, -2.6) {$S^{d+1}$};

  \draw[thick, ->] (0.3, -2) -- (0.3, 2);
  \draw[thick, gray] (0.0, 1.6) -- (7.2, 1.6);
  \node[left, font=\small] at (0.0, 1.6) {$\eta\!=\!0$};

  \coordinate (P1) at (1.0, 1.6);
  \coordinate (P2) at (2.5, 1.6);
  \coordinate (P3) at (4.2, 1.6);
  \coordinate (P4) at (5.8, 1.6);

  \coordinate (W1) at (1.75, -1.4);
  \coordinate (W2) at (4.75, -1.4);

  \draw[thick] (W1) to[bend left=35]  (W2); 
  \draw[thick] (W1) to[bend left=18]   (W2); 
  \node[font=\small] at (3.25, -1.45) {$\vdots$};
  \draw[thick] (W1) to[bend right=35] (W2);  
 
  \draw[thick] (W1) -- (P1);
  \draw[thick] (W1) -- (P2);
  \draw[thick] (W2) -- (P3);
  \draw[thick] (W2) -- (P4);

  \foreach \P in {P1, P2, P3, P4}{
    \filldraw[red] (\P) circle (2pt);
  }

  \filldraw[black] (W1) circle (2.5pt);
  \filldraw[black] (W2) circle (2.5pt);

  \node[above, font=\small] at (P1) {$\vec y_1$};
  \node[above, font=\small] at (P2) {$\vec y_2$};
  \node[above, font=\small] at (P3) {$\vec y_3$};
  \node[above, font=\small] at (P4) {$\vec y_4$};

  \node[font=\small, gray] at (3.5, -2.6) {$\mathrm{dS}_{d+1}$};

\end{tikzpicture}
    \caption{An $s$-channel four-point candy diagram with $n\ge 2$ internal lines on the sphere and in dS.}
    \label{fig:candy}
\end{figure}
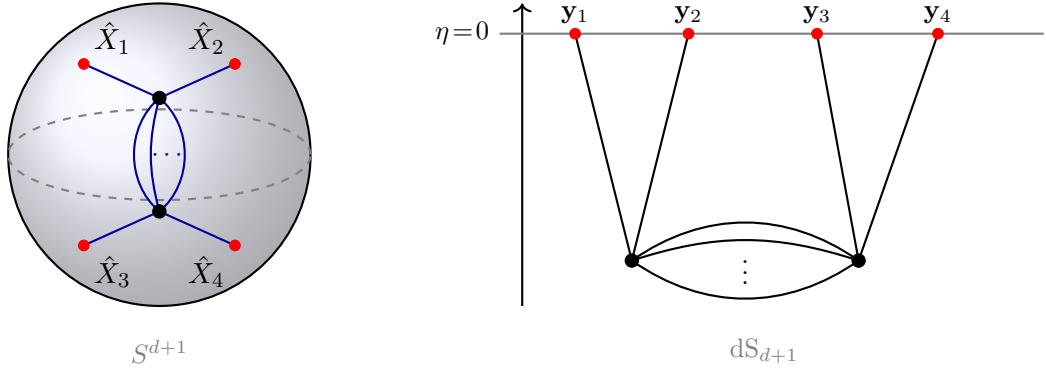

We proceed to verify the equivalence between the two formalisms in the case of candy diagrams, see Figure \ref{fig:candy}. The external and internal lines are associated with scaling dimensions $\Delta_i$ and $\delta_a$ respectively. We focus on the one-loop candy diagram, i.e. $a=1, 2$, and the proof can be generalized to arbitrarily high loop order.

Our strategy is the following. Since the equivalence has already been established for exchange diagrams, we reduce the one-loop bubble in the middle to an effective single internal line. This reduction is implemented by the K\"all\'en--Lehmann (KL) representation on the sphere \cite{Bros:2009bz,Marolf:2010zp,Loparco:2023rug}, which states that the product of two sphere propagators can be decomposed into an integral over a single propagator along the principal series:
\begin{equation}\label{KLexp}
    G_{\delta_1}(\hat X, \hat Y)G_{\delta_2}(\hat X, \hat Y) = \int_{\frac{d}{2}+i\mathbb R}\frac{d\Delta}{2\pi i } \mathfrak{R}_{\delta_1, \delta_2}(\Delta) G_\Delta(\hat X, \hat Y)\,,
\end{equation}
where 
\small
\begin{equation}
    \mathfrak{R}_{\delta_1, \delta_2}(\Delta) = \frac{\Gamma(\frac{\delta_1+\delta_2-\Delta}{2})\Gamma(\frac{\delta_1+\delta_2-\bar\Delta}{2})\Gamma(\frac{\delta_1+\bar\delta_2-\Delta}{2})\Gamma(\frac{\delta_1+\bar\delta_2-\bar\Delta}{2})\Gamma(\frac{\bar\delta_1+\delta_2-\Delta}{2})\Gamma(\frac{\bar\delta_1+\delta_2-\bar\Delta}{2})\Gamma(\frac{\bar\delta_1+\bar\delta_2-\Delta}{2})\Gamma(\frac{\bar\delta_1+\bar\delta_2-\bar\Delta}{2})}{16\pi^{\frac{\bD}{2}}\Gamma(\frac{d}{2})\Gamma(\Delta)\Gamma(\bar\Delta)\Gamma(\Delta-\frac{d}{2})\Gamma(\bar\Delta-\frac{d}{2})}\,.
\end{equation}
\normalsize
For simplicity, we assume throughout the derivation that both $\delta_1$ and $\delta_2$ belong to the principal series. If instead $\delta_1$ and $\delta_2$ lie in the complementary series, a finite number of additional complementary-series propagators may appear on the right-hand side of \eqref{KLexp} \cite{Bros:2009bz,Loparco:2023rug}. But these additional terms do not change our final result, as argued in Footnote \ref{pole}.

The candy diagram can thus be realized as an integral transformation of the exchange diagram. For a fixed exchange diagram, there are two contributions corresponding to the two Witten diagrams, and these contributions are related by  swapping $\Delta$ and $\bar\Delta$. Because of the $\Delta\leftrightarrow\bar\Delta$ symmetry of the KL density $\mathfrak{R}_{\delta_1, \delta_2}(\Delta) $, it suffices to keep only one Witten diagram and double the result
\begin{equation}
    \begin{aligned}
        &\langle \CO_{\Delta_1}(\vec y_1)\cdots \CO_{\Delta_4}(\vec y_4)\rangle^{\SphereAffil{}}_{\rm candy} \\ &=\left(\prod_ic_{\Delta_i}\right)\int_{\frac{d}{2}+i\mathbb R}\frac{d\Delta}{2\pi i } \widetilde{\mathfrak{R}}_{\delta_1, \delta_2}(\Delta)\sin\left(\frac{\Delta_1+\Delta_2-\bar\Delta}{2}\pi\right)\sin\left(\frac{\Delta_3+\Delta_4-\bar\Delta}{2}\pi\right) W^{\rm AdS}_{\rm ex, \Delta}(\vec y_i)\,,
    \end{aligned}
\end{equation}
where $\widetilde{\mathfrak{R}}_{\delta_1, \delta_2}(\Delta) \equiv 8 c_{\Delta}\mathfrak{R}_{\delta_1, \delta_2}(\Delta)$, and $W^{\rm AdS}_{\rm ex, \Delta}(\vec y_i)$ denotes the exchange Witten diagram. 

We then close the contour on the right. There are single poles at \footnote{For $\delta_1$ and $\delta_2$ belonging to complementary series, some of these poles may lie on the left of the integral contour. The missing poles are exactly compensated by the additional complementary-series contribution to the KL decomposition \eqref{KLexp}. \label{pole}}
\begin{equation}
    \delta_1+\delta_2+2n, \quad \delta_1+\bar\delta_2+2n, \quad \bar\delta_1+\delta_2+2n, \quad \bar\delta_1+\bar\delta_2+2n,\quad n\in\mathbb{Z}_{\geq0}\,.
\end{equation}
Focusing on the first set of poles, the residues of $\tilde{\mathfrak{R}}_{\delta_1, \delta_2}(\Delta)$ at $\delta_1+\delta_2+2n$ appear to  satisfy 
\begin{equation}\label{KLtoKL}
    {\rm Res}_{\Delta=\delta_1+\delta_2+2n}\,   \widetilde{\mathfrak{R}}_{\delta_1, \delta_2}(\Delta) = - 4\, c_{\delta_1}c_{\delta_2} a^{\rm AdS}_{\delta_1,\delta_2}(n).
\end{equation}
$ a^{AdS}_{\delta_1,\delta_2}(n)$ are KL coefficients in EAdS, satisfying $\Pi_{\delta_1}\Pi_{\delta_2} = \sum_{n\ge 0}  a^{\rm AdS}_{\delta_1,\delta_2}(n) \Pi_{\delta_1+\delta_2+2n}$, where  $\Pi_\delta$ denotes the free EAdS propagator of boundary dimension $\delta$. The explicit expression of the KL coefficients was calculated in  \cite{Fitzpatrick:2011hu}.
Summing up all the poles, we obtain 
\begin{equation}\label{candyfull}
    \begin{aligned}
        &\langle \CO_{\Delta_1}(\vec y_1)\cdots \CO_{\Delta_4}(\vec y_4)\rangle^{\SphereAffil{}}_{\rm candy}=4 \left(\prod_{i}c_{\Delta_i}\right)\\
        & \times\bigg(c_{\delta_1}c_{\delta_2} \sin\left(\frac{\Delta_1\!+\!\Delta_2\!+\!\delta_1\!+\!\delta_2\!-\!d}{2}\pi\right)\sin\left(\frac{\Delta_3\!+\!\Delta_4\!+\!\delta_1\!+\!\delta_2\!-\!d}{2}\pi\right) W^{\rm AdS}_{\delta_1, \delta_2}(\vec y_i) \\
        &+c_{\bar\delta_1}c_{\delta_2} \sin\left(\frac{\Delta_1\!+\!\Delta_2\!+\!\bar\delta_1\!+\!\delta_2\!-\!d}{2}\pi\right)\sin\left(\frac{\Delta_3\!+\!\Delta_4\!+\!\bar\delta_1\!+\!\delta_2\!-\!d}{2}\pi\right) W^{\rm AdS}_{\bar\delta_1, \delta_2}(\vec y_i) \\
        &+c_{\delta_1}c_{\bar\delta_2} \sin\left(\frac{\Delta_1\!+\!\Delta_2\!+\!\delta_1\!+\!\bar\delta_2\!-\!d}{2}\pi\right)\sin\left(\frac{\Delta_3\!+\!\Delta_4\!+\!\delta_1\!+\!\bar\delta_2\!-\!d}{2}\pi\right) W^{\rm AdS}_{\delta_1, \bar\delta_2}(\vec y_i) \\
        &+c_{\bar\delta_1}c_{\bar\delta_2} \sin\left(\frac{\Delta_1\!+\!\Delta_2\!+\!\bar\delta_1\!+\!\bar\delta_2\!-\!d}{2}\pi\right)\sin\left(\frac{\Delta_3\!+\!\Delta_4\!+\!\bar\delta_1\!+\!\bar\delta_2\!-\!d}{2}\pi\right) W^{\rm AdS}_{\bar\delta_1, \bar\delta_2}(\vec y_i) \bigg)\,,
    \end{aligned}
\end{equation}
where $W^{\rm AdS}_{\delta_1, \delta_2}(\vec y_i)$ denotes the EAdS candy diagram with boundary dimensions $\delta_1,\delta_2$ for the two internal lines. The RHS of \eqref{candyfull} precisely reproduces the in-in calculation of the one-loop candy diagram \cite{Sleight:2021plv}. This method can be applied inductively to show the equivalence for arbitrary higher-loop candy diagrams.

\section{Exchange diagrams: from position space to momentum space}\label{sec:exch}
Starting from the Euclidean sphere, we have derived the Euler integral representation~\eqref{ss} for the exchange contribution to cosmological correlators. We define the corresponding momentum-space correlators by
\begin{equation}\label{Fdef}
    \int_{\vec y} e^{i \sum_i  \vec k_i \cdot \vec y_i}\left\langle \CO_{\Delta_1}(\vec y_1)\cdots\CO_{\Delta_4}(\vec y_4)\right\rangle_{\rm ex}^{\SphereAffil{}}= (2\pi)^d \delta^d\left(\sum_i \vec k_i\right) \CF_{\{\Delta_i\}}^{\Delta} (\vec k_i)\,.
\end{equation}
In this section, starting with the Euler integral representation, we will first show how to extract the collapsed limit, and then derive a MB integral representation which was obtained in \cite{Sleight:2019mgd} using a completely different approach.

Carrying out the Fourier transformation in \eqref{Fdef} leads to 
\begin{equation}\label{tI}
    \CF_{\{\Delta_i\}}^{\Delta} (\vec k_i) = \frac{\CC_{\rm ex}}{2^7 \pi^{3} }\int_{\rho, \sigma,\mu_i}\frac{\rho^{\bar\Delta-\frac{d}{2}}\sigma^{\Delta-\frac{d}{2}}\prod_i\mu_i^{\Delta_i-\frac{d}{2}} }{(1\!+\!\rho^2\!+\!\sigma^2)^{\frac{\bD-\Delta_{\rm tot}}{2}}P_L^{\frac{d}{2}}P_R^{\frac{d}{2}}} e^{-\frac{1}{4P_R}\left(\frac{k_1^2}{\mu_1}+\frac{k_2^2}{\mu_2}\right)- \frac{1}{4P_L}\left(\frac{k_3^2}{\mu_3}+\frac{k_4^2}{\mu_4}\right)-\frac{1\!+\!\rho^2\!+\!\sigma^2}{4\rho\sigma P_L P_R} k_I^2}\,,
\end{equation}
where $k_i\equiv |\vec k_i|$ is the magnitude of the external momenta, $k_I \equiv |\vec k_1+\vec k_2|$ is the magnitude of the exchanged momentum, and the functions $P_L$ and $P_R$ are given  by
\begin{equation}\label{Ps}
    P_L = (\mu_1+\mu_2)\rho\sigma+(\mu_3+\mu_4)(1+\rho^2), \quad P_R = (\mu_3+\mu_4)\rho\sigma+(\mu_1+\mu_2)(1+\sigma^2)\,.
\end{equation}
Following the convention of \cite{Arkani-Hamed:2018kmz}, we define $u = \frac{k_I}{k_{12}}$ and $v=\frac{k_I}{k_{34}}$, where $k_{12} \equiv k_1+k_2$ and $k_{34}\equiv k_3+k_4$. The collapsed limit corresponds to $k_I\to 0$ while all $k_i$ remain finite. In this limit, $u, v\to 0$. For later convenience, we also define $\mu_{12}\equiv \mu_1+\mu_2$ and $\mu_{34} \equiv \mu_3+\mu_4$.

\subsection{The collapsed limit}
As shown in \cite{Arkani-Hamed:2015bza}, the exchange in-in diagram in dS$_4$ with conformally coupled external scalars and a massive intermediate particle of scaling dimension $\frac{3}{2}+i\nu$ exhibits two leading non-analytic behaviors of the same order, $k_I^{\pm 2i\nu}$ in the collapsed limit. This feature also holds in arbitrary spacetime dimensions and for arbitrary external particles \cite{Sleight:2019mgd}. When the intermediate particle belongs to the complementary series, corresponding to  $\nu\to i\nu$ with $\nu\in (0,3/2)$, the leading non-analytic behavior becomes $k_I^{-2\nu}$ \cite{Arkani-Hamed:2018kmz}. In this subsection, we carry out the derivation for the complementary series, with the final result obtained for the principal series by analytic continuation.

In the Schwinger parameter integral \eqref{tI}, we can  obtain these nonanalytic behaviors by zooming into the small $\rho$ or small $\sigma$ regions in the collapsed limit. To illustrate the idea, we consider a toy example of this integral, namely the integral representation of Bessel $K$ function:
\begin{equation}\label{BK}
    \int_0^\infty\frac{dx}{x^{1+\alpha}} e^{-x - \frac{t^2}{4x}}= 2^{1+\alpha} t^{-\alpha} K_\alpha(t) = \left(\Gamma(-\alpha)+O(t)\right)+ t^{-2\alpha}\left(4^\alpha \Gamma(\alpha)+O(t)\right)\,,
\end{equation}
where in the last step, we have used the series expansion of Bessel functions. There are two fall-offs at $t=0$, with $t^0$ dominating for negative $\alpha$ and $t^{-2\alpha}$ dominating for positive $\alpha$. We can obtain these leading fall-offs easily without knowing anything about the Bessel function. Setting $t=0$ in the exponent, the integral is convergent and yields $\Gamma(-\alpha)$ when $\alpha<0$. This corresponds to the $\Gamma(-\alpha) t^0$ term in \eqref{BK}. On the other hand, substitute  $x\to t^2 x$, and drop the $t^2 x$ term in the exponent. The resulting  integral is convergent and yields $(2/t)^{2\alpha}\Gamma(\alpha)$ when $\alpha>0$, which is precisely the second term in \eqref{BK}. 

Inspired by  this idea and noticing the factor $\frac{1+\rho^2+\sigma^2}{ P_L P_R} \frac{k_I^2}{4\rho\sigma}$ in the exponent of \eqref{tI} we make the rescaling  $\sigma\to k_I^2\sigma$.\footnote{Physically, the non-analytic behavior in the collapsed limit is captured by the leading long distance behavior of the internal line, which is also equivalent to the late time limit. As we have shown before, rescaling $\sigma$ (or $\rho$) is exactly the way to extract the late time limit.} The $\sigma^{\Delta-\frac{d}{2}}$ term gives the expected power $k_I^{-2\nu}$ for $\Delta =\frac{d}{2}-\nu$. In the rest part of the integrand, we take the $k_I\to 0$ limit. So $P_L$ reduces to $\mu_{34}(1+\rho^2)$, and $P_R$ reduces to $\mu_{12}$. The $\sigma$ integral becomes $\int_\sigma \sigma^{-\nu} \exp(-\frac{1}{4\rho \sigma\mu_{12}\mu_{34}})=\Gamma(\nu)(4\rho\mu_{12}\mu_{34})^{\nu}$. The $\rho$ integral is of the form $\int_\rho \rho^{\#1}/(1+\rho^2)^{\#2}$ after  performing another rescaling $(\mu_3,\mu_4)\to 1/\sqrt{1+\rho^2}(\mu_3,\mu_4)$. There is one more simplification in that we can identify $k_1$ with $k_2$ and $k_3$ with $k_4$ in the collapsed limit. Altogether, the remaining $\mu_i$ integrals factorize into two simple twodimensional integrals
\begin{equation}
    \left.\CF^{\Delta=d/2-\nu}_{\{\Delta_i\}}(\vec k_i)\right|_{k_I\to 0} =  \frac{2^{-8}\pi^{-3}\Gamma(\frac{d}{2}\!-\!\Delta)^2\prod_i \Gamma(\frac{d}{2}\!-\!\Delta_i)}{ \Gamma(\frac{\bar\Delta\!-\!\Delta_1\!-\!\Delta_2}{2}\!+\!1)\Gamma(\frac{\bar\Delta\!-\!\Delta_3\!-\!\Delta_4}{2}\!+\!1)}\left(\frac{k_I^2}{4}\right)^{\Delta-d/2}h^{\Delta}_{\Delta_1, \Delta_2}(k_1)h^{\Delta}_{\Delta_3, \Delta_4}(k_3)~,
\end{equation}
where $h^{\Delta}_{\Delta_1, \Delta_2}(k_1)$ (and similarly $h^{\Delta}_{\Delta_3, \Delta_4}(k_3)$) is given by 
\begin{equation}
    \begin{aligned}
        &h^{\Delta}_{\Delta_1, \Delta_2}(k_1)= \int_{\mu_1, \mu_2} \frac{\mu_1^{\Delta_1-d/2}\mu_2^{\Delta_2-d/2}}{(\mu_1+\mu_2)^\Delta} e^{-\frac{k_1^2}{4\mu_1\mu_2}}\\
        &=\frac{\Gamma(\frac{\bar\Delta_1+\bar\Delta_2-\bar\Delta}{2})\Gamma(\frac{\Delta+\Delta_1-\Delta_2}{2})\Gamma(\frac{\Delta+\Delta_2-\Delta_1}{2})}{2\Gamma(\Delta)}\left(\frac{k_1}{2}\right)^{\bar\Delta-\bar\Delta_1-\bar\Delta_2}\,.
    \end{aligned}
\end{equation}

For $\Delta=\frac{d}{2}+i\nu$, there is another contribution proportional to $k_I^{d-2\Delta}$ at this order, which is obtained by zooming into the small $\rho $ region. Combining the two contributions, we get
\begin{equation}\label{OPEgen}
\begin{aligned}
    &\left.\CF^{\Delta=\frac{d}{2}+i\nu}_{\{\Delta_i\}}(\vec k_i)\right|_{k_I\to 0} = \frac{2^{3d-\Delta_{\rm tot}}\prod_i \Gamma(\frac{d}{2}-\Delta_i)}{(4\pi)^5k_1^{\bar\Delta_1+\bar\Delta_2-\frac{d}{2}}k_3^{\bar\Delta_3+\bar\Delta_4-\frac{d}{2}}}\left[\frac{\Gamma(-i\nu)^2}{\Gamma(\Delta)^2}\sin\left(\frac{\Delta_1\!+\!\Delta_2\!-\!\bar\Delta}{2}\pi\right)\sin\left(\frac{\Delta_3\!+\!\Delta_4\!-\!\bar\Delta}{2}\pi\right)\right.\\
    &\left.\times\Gamma\left(\frac{\Delta\pm(\Delta_1\!-\!\frac{d}{2})\hat\pm(\Delta_2\!-\!\frac{d}{2})}{2}\right)\Gamma\left(\frac{\Delta\pm(\Delta_3\!-\!\frac{d}{2})\hat\pm(\Delta_4\!-\!\frac{d}{2})}{2}\right)\left(\frac{k_I^2}{k_1 k_3}\right)^{i\nu}  + (\Delta\leftrightarrow\bar\Delta)\right]\,,
\end{aligned}
\end{equation}
where we have used the shorthand notation $\Gamma(a\pm b)\equiv \Gamma(a+b)\Gamma(a-b) $. Eq. \eqref{OPEgen} is in precise agreement with \cite{Sleight:2019mgd} upon the replacement $k_1\sim \frac{1}{2}k_{12}$ and $k_3\sim \frac{1}{2}k_{34}$. In ${\rm dS}_4$, for external particles with $\Delta_i=1$, \eqref{OPEgen} agrees with the result of \cite{Arkani-Hamed:2015bza}, namely
\begin{equation}
  {\rm dS}_4:\quad\left.\mathcal{F}^{\Delta=\frac{3}{2}+i\nu}_{\{\Delta_i=1\}}(\vec k_i)\right |_{k_I\to 0}=\frac{1}{4}\left[\frac{\left(1\!+\!i\sinh(\pi\nu)\right)\Gamma\left(\frac{1}{2}\!+\!i\nu\right)^2\Gamma(-i\nu)^2}{8\pi k_1 k_2 k_3 k_4 \sqrt{k_{12}k_{34}}}\left(\frac{uv}{4}\right)^{i\nu}+(\nu\leftrightarrow-\nu)\right]\, .
\end{equation}
The remaining overall factor of $1/4$ is due to a difference in interaction normalization. In \cite{Arkani-Hamed:2015bza}, the cubic interaction is chosen as $\int \lambda\varphi^2\sigma$, which gives an additional combinatorial factor $2\times 2=4$ for the four-point  exchange diagram of $\varphi$.

\subsection{Mellin representation}
To derive the full expression of $\CF$, including analytic terms in $k_I$, we will derive a Mellin representation of \eqref{tI}. For simplicity of the derivation, we focus on the case of conformally  coupled external legs, i.e. $\Delta_i = \frac{d-1}{2}$.

Notice that most terms in \eqref{tI} depend on the sum $\mu_{12}=\mu_1+\mu_2$. So, we first integrate out $t_L\equiv \frac{\mu_1}{\mu_{12}}\in [0,1]$. The relevant integral is 
\begin{equation}
    \int_0^1\frac{dt_L}{(t_L(1-t_L))^{\frac{3}{2}}}\exp\left(-\frac{1}{4\mu_{12}P_R}\left(\frac{k_1^2}{t_L}+\frac{k_2^2}{1-t_L}\right)\right)=2\sqrt{\pi\mu_{12}P_R}\frac{k_{12}}{k_1k_2}\exp\left(-\frac{k_{12}^2}{4\mu_{12}P_R}\right)\,.
\end{equation}
This simple result is specific to conformally coupled scalars. For generic $\Delta_i$, the $t_L$ integral cannot be evaluated analytically and one may trade the $t_L$ integral for a Mellin-Barnes integral. Similarly, we integrate out $t_R= \frac{\mu_3}{\mu_{34}}$. Then we make the substitution $\mu_{34} = \tau\mu_{12}$ and integrate out $\mu_{12}$, which is a standard Schwinger parameter integral. The remaining $\rho,\sigma$ and $\tau$ integral becomes 
\begin{equation}
    \CF_{\{\Delta_i=\frac{d-1}{2}\}}^{\Delta=\frac{d}{2}+i\nu}(\vec k_i)=\frac{2^{d-6}\CC_{\rm ex} k_{12}k_{34}}{\pi^2 k_1k_2k_3 k_4}\int_{\rho,\sigma,\tau}\frac{\rho^{-i\nu}\sigma^{i\nu}}{(1+\rho^2+\sigma^2)^{2-\frac{d}{2}}}\frac{\Gamma(\frac{d}{2})\sqrt{p_Lp_R}}{\left(k_{12}^2 p_L +k_{34}^2 p_R+\frac{1+\rho^2+\sigma^2}{\rho\sigma}k_I^2\right)^{\frac{d}{2}}}~,
\end{equation}
where $p_L=P_L/\mu_{12} = \rho\sigma+(1+\rho^2)\tau$ and $p_R=P_R/\mu_{34}=\rho\sigma+(1+\sigma^2)/\tau$. To proceed, we introduce two Mellin-Barnes integrals 
\begin{equation}
    \CF_{\{\Delta_i=\frac{d-1}{2}\}}^{\Delta=\frac{d}{2}+i\nu}(\vec k_i)=\frac{2^{d-6}\CC_{\rm ex} k_{12}k_{34} }{\pi^2 k_1k_2k_3 k_4}\int\frac{dzdw}{(2\pi i)^2}\frac{\Gamma(-z)\Gamma(-w)\Gamma(\frac{d}{2}\!+\!z\!+\!w)}{k_I^d \,u^{2z}v^{2w}}\int_{\rho,\sigma,\tau}\frac{\rho^{\bar\Delta+z+w}\sigma^{\Delta+z+w} p_L^{\frac{1}{2}+z}p_R^{\frac{1}{2}+w}}{(1+\rho^2+\sigma^2)^{2+z+w}}~.
\end{equation}
The Mellin-Barnes contours are straight lines extending along the imaginary direction and  satisfying  $\Re(z)<0, \Re(w)<0$ and $\Re(z+w)>-\frac{d}{2}$. Integrating out $\tau$ yields two hypergeometric functions. After expanding them into series, the $\rho$ and $\sigma$ integrals appear to be decoupled:
\begin{equation}
    \int_\tau \frac{p_L^{\frac{1}{2}+z}p_R^{\frac{1}{2}+w}}{(1\!+\!\rho^2\!+\!\sigma^2)^{2+z+w}} =\sum_{n\ge 0} \frac{ B(-z\!-\!\frac{1}{2},z\!-\!w)
   \left(w+\frac{3}{2}\right)_n^2(\rho\sigma) ^{2n+w-z}}{n! (w-z+1)_n \left(1+\rho
   ^2\right)^{\frac{3}{2}+w+n}\left(1+\sigma
   ^2\right)^{\frac{3}{2}+w+n}}+(z\leftrightarrow w)\,,
\end{equation}
where $B(a, b)\equiv \frac{\Gamma(a)\Gamma(b)}{\Gamma(a+b)}$ is the Euler beta function. The $\tau$ integral is convergent when $\Re(z), \Re(w)<-\frac{1}{2}$.

For each $n$, the integrations  over $\rho$ and $\sigma$ are straightforward. The convergence condition, for all $n\ge 0$, imposes $d<6$ and $\Re(z), \Re(w)>-\frac{d}{4}$ \footnote{Here we have assumed that the internal line belongs to the principal series. Otherwise, the constraint is $\bar\Delta<3$ and $\Re(z),\Re(w)>-\frac{\Delta}{2}$, in our convention $\Delta\le\frac{d}{2}\le\bar\Delta$.}. The remaining sum over $n$ leads to hypergeometric functions with the last argument being 1 and hence reduces to some $\Gamma$ functions, provided $\Re(z+w)< 1-\frac{d}{2}$. Altogether, after the contour shift $z\to z-\frac{d}{4}$ and $w\to w-\frac{d}{4}$, we obtain the following double MB representation of the exchange diagram 
\small
\begin{equation}\label{MBfinal}
\begin{aligned}
    &\CF_{\{\Delta_i=\frac{d-1}{2}\}}^{\Delta=\frac{d}{2}+i\nu}(\vec k_i)= \frac{k_I^{2-d}(u v)^{\frac{d}{2}-1}}{2^5\pi k_1k_2k_3k_4} \int\frac{dzdw}{(2\pi i)^2}\frac{4^{z+w}\kappa(z, w)}{u^{2z}v^{2w}} \Gamma\left(z\!\pm\! \frac{i\nu}{2}\right)\Gamma\left(w\!\pm\! \frac{i\nu}{2}\right)\Gamma\left(\frac{d\!-\!2}{2}\!-\!2z\right)\Gamma\left(\frac{d\!-\!2}{2}\!-\!2w\right)\,,\\
    &\kappa(z, w) \equiv \frac{\cos ^2\left(\pi  \left(w-\frac{d}{4}\right)\right) (\cos (2 \pi  z)-\cosh (\pi  \nu ))-\cos^2\left(\pi  \left(z-\frac{d}{4}\right)\right) (\cos (2 \pi  w)-\cosh (\pi  \nu ))}{\sin(\pi(w-z))\sin(\pi(w+z))}\,,
\end{aligned}
\end{equation}
\normalsize
with the contours being $z\in c_1+i\bR$ and $w\in c_2+i\bR$, where $c_1$ and $c_2$ satisfy 
\begin{equation}
    {\rm Contour}: \quad 0<c_1, c_2<\frac{d-2}{4}, \quad c_1+c_2<1.
\end{equation}
In the $z$-plane, the poles of $\Gamma(z\pm\frac{i\nu}{2})$ are on the left of the $z$-contour, and the poles of $\Gamma(\frac{d-2}{2}-2z)$ are on the right of the $z$-contour. $\kappa(z, w)$ has poles at $z=z_n\equiv-w+n$. $z_n$ with $n\ge 1$ lie on the right of the $z$-contour and $z_n$ with $n\le 0$ lie on the left of the $z$-contour. Eq. \eqref{MBfinal} appears to match the Mellin-Barnes representation obtained in \cite{Sleight:2019mgd}. The derivation in \cite{Sleight:2019mgd} is based on a Mellin space representation of de Sitter propagators, and the bulk integrals follow the in-in contour. Surprisingly, we introduce the Mellin variables for a completely different purpose but arrive at the same results.

With the MB representation, we can derive the EFT expansion which is analytic in $k_I$. Assuming $u<v$, we first close the  $z$-contour to the left, picking up poles at  $z=-w-m$, $m\in\mathbb{Z}_{\geq0}$. We then close the remaining $w$-contour to the right, selecting poles at $w=\frac{d-2}{4}+\frac{n}{2}$, $n\in\mathbb{Z}_{\geq0}$. Thus, we can obtain a Taylor expansion in the two variables $u/v$ and $u$, known as the EFT expansion \cite{Arkani-Hamed:2018kmz, Sleight:2019mgd}.

\section{Landau analysis of singularities}\label{sec:Landau}

As for scattering amplitudes, the singularity structure of cosmological correlators is an important ingredient in determining them. The bootstrap of tree-level cosmological correlators, for instance, relies heavily on knowledge of their singularities. In this section we show that Schwinger-parameter integrals provide a convenient and systematic way to locate singularities in momentum space without evaluating the integrals explicitly. The strategy is to apply Landau analysis to these integrals\footnote{For a comprehensive review on Landau analysis, see \cite{Eden:1966dnq}. For its earlier application in cosmology, see \cite{Salcedo:2022aal,Pimentel:2026kqc}}. Consider, for illustration, a single-variable integral that depends on external parameters. At special values of the parameters, either a singularity of the integrand approaches an endpoint of the integration domain, or two singularities of the integrand pinch the integration contour. The integral can be singular only at such points in parameter space.
To make the idea more concrete, we study a toy integral
\begin{equation}\label{Iab}
    I(a, b) = \int_0^1dx\frac{\sqrt{x(1-x)}}{\frac{a^2}{x}+\frac{b^2}{1-x}} = \frac{\pi}{16}\frac{a^2+4ab+b^2}{(a+b)^4}~.
\end{equation}
The equality initially holds for positive $a$ and $b$, and the RHS can be analytically continued to the full complex plane, where we find a singularity at $a=-b$. We would like to identify this singularity without explicitly evaluating the integral.

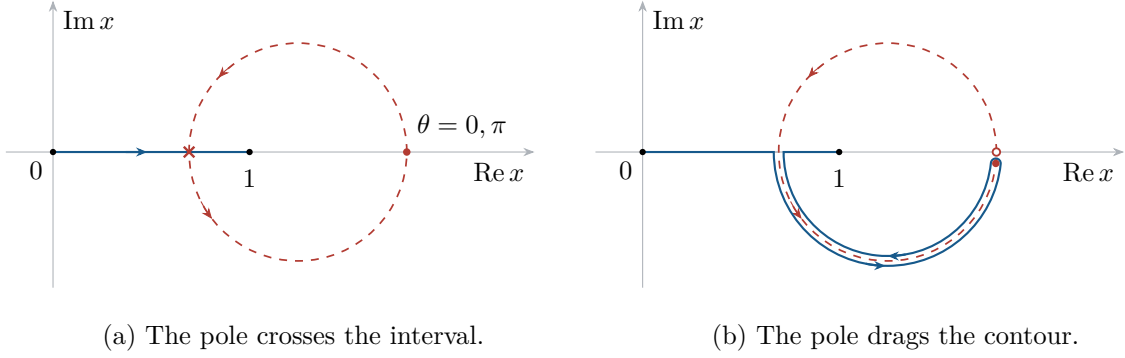
\begin{figure}[t]
    \centering 
\begin{tikzpicture}[
  x=2.6cm,y=2.6cm,font=\small,
  axis/.style={axisgray,thin,-{Stealth[length=1.5mm]}},
  contour/.style={contourblue,line width=.8pt},
  trajectory/.style={polered,dashed,line width=.65pt},
  direction/.style={-{Stealth[length=1.5mm,width=1.2mm]}}
]

\pgfmathsetmacro{\xcross}{9/13}
\pgfmathsetmacro{\cc}{81/65}
\pgfmathsetmacro{\rr}{36/65}
\def\tube{0.025}       
\def\ending{354}       
\pgfmathsetmacro{\px}{\cc+\rr*cos(\ending)}
\pgfmathsetmacro{\py}{\rr*sin(\ending)}

\foreach \shift in {0,3} {
  \begin{scope}[xshift=\shift*2.6cm]
    \draw[axis] (-.24,0)--(2.45,0)
      node[below left,black] {$\operatorname{Re}x$};
    \draw[axis] (0,-.69)--(0,.77)
      node[below right,black] {$\operatorname{Im}x$};
    \draw[trajectory] (\cc,0) circle[radius=\rr];
    \draw[polered,direction]
      ({\cc+\rr*cos(126)},{\rr*sin(126)})
      arc[start angle=126,end angle=137,radius=\rr];
    \draw[polered,direction]
      ({\cc+\rr*cos(207)},{\rr*sin(207)})
      arc[start angle=207,end angle=218,radius=\rr];
  \end{scope}
}

\node[anchor=west] at (0.2,-.94) {(a) The pole crosses the interval.};
\draw[contour] (0,0)--(1,0);
\draw[contour,direction] (.37,0)--(.48,0);
\fill (0,0) circle[radius=.016] node[below left] {$0$};
\fill (1,0) circle[radius=.016] node[below=3pt] {$1$};
\fill[polered] (1.8,0) circle[radius=.019];
\node[anchor=west] at (1.80,.13) {$\theta=0,\pi$};
\draw[polered,line width=.9pt]
  ({\xcross-.027},-.027)--({\xcross+.027},.027)
  ({\xcross-.027},.027)--({\xcross+.027},-.027);
\begin{scope}[xshift=7.8cm]
  \node[anchor=west] at (0.3,-.94)
    {(b) The pole drags the contour.};
  \draw[axisgray,dashed] (0,0)--(1,0);
  \draw[contour]
    (0,0)--({\cc-\rr-\tube},0)
    arc[start angle=180,end angle=\ending,radius={\rr+\tube}]

    arc[start angle=\ending,end angle={\ending+180},radius=\tube]

    arc[start angle=\ending,end angle=180,radius={\rr-\tube}]
    --(1,0);
  \draw[contour,direction]
    ({\cc+(\rr+\tube)*cos(259)},{(\rr+\tube)*sin(259)})
    arc[start angle=259,end angle=269,radius={\rr+\tube}];
  \draw[contour,direction]
    ({\cc+(\rr-\tube)*cos(280)},{(\rr-\tube)*sin(280)})
    arc[start angle=280,end angle=270,radius={\rr-\tube}];
  \fill (0,0) circle[radius=.016] node[below left] {$0$};
  \fill (1,0) circle[radius=.016] node[below=3pt] {$1$};
  \draw[polered,fill=white,line width=.7pt]
    (1.8,0) circle[radius=.019];
  \fill[polered] (\px,\py) circle[radius=.019];
\end{scope}
\end{tikzpicture}
   \caption{Analytic continuation of $b = q\,e^{i\theta}$ in the integral $I(a, b)$. Here, we take $a>q>0$.}
    \label{Landauillu}
\end{figure}

First, we notice that the integrand has a pole at $x_*=\frac{a^2}{a^2-b^2}$, which lies outside the integration interval $x\in[0,1]$ for $a,b>0$. Keeping $a>0$ fixed, we analytically continue $b=qe^{i\theta}$, with $0<q<a$ and $0\leq\theta\leq\pi$. The trajectory of $x_*(\theta)$ is shown by the dashed line in panel (a) of Figure \ref{Landauillu}. At $\theta=\pi/2$, the pole crosses the original contour from the upper to the lower half-plane. To continue $I(a,b)$ beyond this point, we deform the contour below the moving pole. Panel (b) illustrates the resulting contour for $\theta$ close to $\pi$. We now take $q\to a^{-}$, so that $b=-q\to-a$ at $\theta=\pi$. The moving pole $x_*=a^2/(a^2-q^2)$ escapes to infinity, dragging the contour with it. Its collision with the fixed singularity at $x=\infty$ pinches the deformed contour, producing the singularity at $a+b=0$. The degree of this singularity can be obtained by examining the large-$x$ region. Setting $b=-a+\epsilon$ and rescaling $x=au/\epsilon$, we find that the contribution from this region scales as $a^2/\epsilon^4$, in agreement with the exact result \eqref{Iab}.

We can recover this singularity by using the Landau equations. For simplicity we perform a change of variable $x=\frac{y}{1+y}$, which brings the pole at $\infty$ to $L_1: y+1=0$. The other pole is at $L_2: a^2+b^2y=0$. The collision of poles means that $L_1$ and $L_2$ vanish simultaneously. In higher dimensions we also require the singular surfaces have parallel normal directions but in $d=1$ this condition is trivially satisfied. The possible singularities are $a\pm b=0$. It is clear from the integral that $a=b$ is not a singularity, and hence the only singularity is $a=-b$.

\subsection{Contact diagram}
The $n-$point contact diagram in momentum space reads
\begin{equation}\label{DfunctionFourier}
    \mathcal{F}_{\{\Delta_i\}}(\vec k_i)=\pi^{\frac d2(n-1)}\int_{\mu}\frac{\prod_{i=1}^n\mu_i^{\Delta_i-\frac d2}}{\mu_{\rm tot}^{\frac d2(n-2)}}\exp\left(-\frac{1}{4\mu_{\rm tot}}\sum_{i=1}^n\frac{k_i^2}{\mu_i}\right),\qquad \mu_{\rm tot}=\sum_{i=1}^n\mu_i\,.
\end{equation}
Putting $\mu_i=rt_i$ with $\sum_i t_i=1$, the $r$ integral yields a negative power of
$\sum_i k_i^2/t_i$. As in the toy example, we make the substitution
\begin{equation}
    t_a=\frac{u_a}{1+\sum_a u_a},\quad  t_n = \frac{1}{1+\sum_a u_a}~,
\end{equation}
where $a=1, \dots, n-1$. The resulting integral has two singular surfaces:
\begin{equation}
    L_1=\sum_a u_a+1=0, \quad L_2 = \sum_a \frac{k_a^2}{u_a} + k_n^2 =0 ~.
\end{equation}
The Landau condition $dL_1\parallel dL_2$ implies that $k_a^2/u_a^2$ is independent of $a$. We define $u_a = \sigma_a k_a\lambda$, where $\sigma_a\in\{\pm\}$. Plugging this into $L_1=0$ gives $\lambda = -1/\sum_a\sigma_a k_a$. Then the remaining condition $L_2=0$ can be conveniently formulated as $\sum_{i=1}^n \sigma_i k_i=0$, with $\sigma_i\in\{\pm 1\}$.
As shown in the toy example, the continuation of the positive momenta selects $\sigma_i=1$ for every $i$, leading to the total energy singularity $\sum_i k_i=0$. In the Lorentzian picture, this singularity is due to the time integral close to the far past.

\subsection{Tree-level exchange diagram}
In eq.~\eqref{tI}, we use the parameterization $\mu_1=st_L$, $\mu_2=s(1-t_L)$, $\mu_3=s\tau t_R$, and $\mu_4=s\tau(1-t_R)$ and carry out the elementary $s$ integral, leading to
\begin{equation}\label{ExchangeLandauIntegral}
\begin{aligned}
    \CF_{\{\Delta_i\}}^\Delta(\vec k_i)&=\frac{\CC_{\rm ex}\Gamma\left(\frac{3d-\Delta_{\rm tot}}{2}\right)}{2^{8+\Delta_{\rm tot}-3d}\pi^3}\int_{\rho,\sigma,\tau}\frac{\rho^{\bar\Delta-\frac d2}\sigma^{\Delta-\frac d2}\tau^{\frac{\Delta_3+\Delta_4-\Delta_1-\Delta_2}{2}}}{(1+\rho^2+\sigma^2)^{\frac{\bD-\Delta_{\rm tot}}2}}\int_0^1\frac{dt_L}{t_L(1-t_L)}\,\frac{dt_R}{t_R(1-t_R)}\,\\
    &\times t_L^{\Delta_1-\frac d2}(1-t_L)^{\Delta_2-\frac d2}t_R^{\Delta_3-\frac d2}(1-t_R)^{\Delta_4-\frac d2}\frac{(p_Lp_R)^{d-\frac{\Delta_{\rm tot}}2}}{\mathcal G_{\rm ex}^{\frac{3d-\Delta_{\rm tot}}2}}\,,
\end{aligned}
\end{equation}
where $\mathcal G_{\rm ex}$ is given by 
\begin{equation}\label{ExchangeLandauFunction}
\begin{aligned}
    \mathcal G_{\rm ex}&=\left(\frac{k_1^2}{t_L}+\frac{k_2^2}{1-t_L}\right)p_L+\left(\frac{k_3^2}{t_R}+\frac{k_4^2}{1-t_R}\right)p_R+\frac{1+\rho^2+\sigma^2}{\rho\sigma}\,k_I^2\,,\\
    p_L&=\rho\sigma+(1+\rho^2)\tau, \qquad   p_R=\rho\sigma+(1+\sigma^2)/\tau\,.
\end{aligned}
\end{equation}
The $s$ integral is convergent because $\Delta_i\le d/2$ for each external operator.

As we have found in the previous section, the singularity in the squeezed limit $k_I\to 0$ arises from small $\rho$ or small $\sigma$. Similarly, we can find the total energy singularity in the large $\rho$ and $\sigma$ regime. To obtain the partial energy singularities, we solve the Landau equations
\begin{equation}\label{ExchangeLandauEquations}
    \mathcal G_{\rm ex}=\partial_{t_L} \mathcal G_{\rm ex}=\partial_{t_R} \mathcal G_{\rm ex}=\partial_\rho\mathcal G_{\rm ex}=\partial_\sigma\mathcal G_{\rm ex}=\partial_\tau\mathcal G_{\rm ex}=0\,.
\end{equation}
The $t_L$ and $t_R$ derivatives set $t_{L,*}=k_1/k_{12}$ and $t_{R,*}=k_3/k_{34}$. The other derivatives pick the saddle point 
\begin{equation}\label{ExchangeFiniteLandau}
    \rho_*=\sigma_*=\sqrt{\frac{k_I}{k_{12}+k_{34}}},\qquad\tau_*=\frac{k_{34}}{k_{12}}\,.
\end{equation}
At this point, we have $ p_{L}=\frac{k_{34}+k_I}{k_{12}}, p_{R}=\frac{k_{12}+k_I}{k_{34}}$ and $\mathcal G_{{\rm ex}}=2(k_{12}+k_I)(k_{34}+k_I)$. Because $\mathcal G_{\rm ex}$ has a larger power than $p_L$ or $p_R$, the exchange diagram has partial energy singularities at $k_{12}+k_I=0$ and $k_{34}+k_I=0$. In the in-in formalism, they correspond to pushing one of the bulk vertices to the far past.

\subsection{Candy diagram}
The same analysis can be straightforwardly extended to loop diagrams, for example, the candy diagrams with $n$ exchanged particles. In Section \ref{sec:candy}, we have used the K\"all\'en-Lehmann spectral decomposition to reduce a one-loop candy diagram to a tree-level exchange diagram with an additional spectral integral. Here, we  use the Schwinger-parameter representation:
\begin{equation}\label{candygen}
    \langle \CO_{\Delta_1}(\vec y_1)\cdots \CO_{\Delta_4}(\vec y_4)\rangle^{\SphereAffil{}}_{\rm candy}=\mathcal{N}_{\rm candy}\int_{\rho, \sigma,\mu}\frac{\prod_{a=1}^n\rho_a^{\bar\delta_a}\sigma_a^{\delta_a} \prod_{i=1}^4\mu_i^{\Delta_i}}{((1+\bm \rho^2)(1+\bm\sigma^2)-(\bm\rho\cdot\bm\sigma)^2)^{\frac{\bD-\Delta_{\rm tot}}{2}}} e^{-\Phi_{\rm candy} }~,
\end{equation}
where $\bm\rho\equiv (\rho_1,\rho_2,\cdots,\rho_n)$ and $\bm\sigma=(\sigma_1,\sigma_2,\cdots,\sigma_n)$ are associated with the $n$ internal lines, and $\mu_i$ are associated with the four external lines. The normalization factor $\mathcal{N}_{\rm candy}$ is  given by 
\begin{equation}
    \mathcal{N}_{\rm candy}=\frac{\prod_i \Gamma(\frac{\bar\Delta_i-\Delta_i}{2})}{2^4(2\pi^{\bD/2})^{n+2} \Gamma\left(\frac{\bD-\Delta_1-\Delta_2-\sum_a\bar\delta_a}{2}\right)\Gamma\left(\frac{\bD-\Delta_3-\Delta_4-\sum_a\delta_a}{2}\right)}
\end{equation}
and $\Phi_{\rm candy}$ is the same as $\Phi_{\rm ex}$ with the substitution: $\rho^2\to\bm\rho^2, \sigma^2\to\bm\sigma^2$ and $\rho\sigma\to\bm\rho\cdot\bm\sigma$.

After performing a Fourier transformation, we find that the singularity of the integrand is encoded in the following function 
\begin{equation}\label{CandyLandauFunction}
\begin{aligned}
    \mathcal G_{\rm candy}&=\left(\frac{k_1^2}{t_L}+\frac{k_2^2}{1-t_L}\right)(\rho\sigma\cos\theta+\tau(1+\rho^2))+\left(\frac{k_3^2}{t_R}+\frac{k_4^2}{1-t_R}\right)\left(\rho\sigma\cos\theta+\frac{1+\sigma^2}{\tau}\right)\\
    &+\frac{1+\rho^2+\sigma^2+\rho^2\sigma^2\sin^2\theta}{\rho\sigma\cos\theta}k_I^2\,.
\end{aligned}
\end{equation}
Here, $t_L, t_R$ have the same  meaning as in the exchange case, $\rho$($\sigma$) denotes the length of the vector $\bm\rho$($\bm\sigma$), and $\theta$ is the angle between the two vectors. In particular, when $\theta=0$, $  \mathcal G_{\rm candy}$ takes the same form as $  \mathcal G_{\rm ex}$. The Landau equations are similar to the exchange case, besides the new equation $\partial_\theta  \mathcal G_{\rm candy}=0$. Notice that the $\theta$ derivative is proportional to $\sin\theta$, so it has an obvious saddle point at $\theta=0$. When $\theta=0$, the calculation is exactly the same as for the exchange diagram, and we obtain the partial energy singularities. On the other hand, if we strip off the $\sin\theta$ factor in $\partial_\theta  \mathcal G_{\rm candy}$, all the Landau conditions reduce to only three independent equations:
\begin{equation}
    \frac{k_{12}^2+k_{34}^2-k_I^2}{k_I^2} = \frac{(1+\rho^2)(1+\sigma^2)}{\rho^2\sigma^2\cos^2\theta}, \quad k_{12}^2\rho\tau+k_I^2\frac{1+\sigma^2}{\sigma\cos\theta}=0, \quad k_{34}^2\frac{\sigma}{\tau}+k_I^2\frac{1+\rho^2}{\rho\cos\theta}=0\,,
\end{equation}
where we have set $t_L=k_1/k_{12}$ and $t_R=k_3/k_{34}$. For these equations to be self-consistent, the external momenta need to satisfy $(k_I^2-k_{12}^2)(k_I^2-k_{34}^2)=0$, which again implies the partial energy singularities. Therefore, the candy diagrams have the same singularities as the tree-level exchange diagram. This agrees with the explicit result for the one-loop candy diagram with all propagators being conformally coupled \cite{Pimentel:2026kqc}.

Following the same derivation as in Section \ref{sec:exch}, we  also obtain the non-analytic behavior in the collapsed limit. The $k_I$ dependence takes the form $ k_I^{2\sum_a\gamma_a - d}$, where each $\gamma_a$ takes a value in $\{\delta_a, \bar\delta_a\}$. So, there are $2^n$ leading powers.

\section{Conclusions and outlook}
By using integrations on the sphere, we have developed a framework for transforming an arbitrary massive scalar in-in correlator in dS which involves summing up all different in-in contours into a single generalized Euler integral, with one Schwinger parameter associated with each external line and two Schwinger parameters associated with each internal line. The Euler integral representation enables us to explicitly verify the equivalence between the Euclidean formalism and the in-in formalism of dS cosmological correlators for the contact, tree-level exchange and $n$-loop candy diagrams. Furthermore, by applying the Landau singularity equations to the Euler integral representation, we recover the singularities of these diagrams. Below, we list potential extensions and applications of this Euclidean framework of cosmological correlators:

\begin{itemize}
\item \textbf{Systematic study of singularities.} Our analysis of singularities extends naturally to more complicated loop diagrams, such as the triangle and box diagrams contributing to four-point correlators. We expect these to exhibit new types of singularities beyond the examples considered in this paper. Moreover, we have not determined the explicit behavior of the correlators near these singularities, nor the discontinuities across the associated branch cuts. These data would provide input for a bootstrap of cosmological correlators at higher loops.

\item \textbf{Numerical evaluation.} Ref.~\cite{Borinsky:2026roj} applied two numerical methods to closely related Euler integrals\footnote{In their framework, there is an Euler integral for each in-in branch choice of every interacting vertex \cite{Herderschee:2025znl}.}: tropical Monte Carlo integration \cite{Borinsky:2020rqs}, and the reduced Schwinger method introduced there.\footnote{As noted in~\cite{Borinsky:2026roj}, the reduced Schwinger method was developed by the AI model Claude Fable~5.} We expect both methods to carry over to the Euler integrals derived here, providing powerful tools for evaluating cosmological correlators numerically within our framework.

\item \textbf{Inflationary backgrounds.} Our derivation of the Euler integral representation relies on the full $\mathrm{SO}(d+2)$ isometry of the Euclidean sphere $S^{d+1}$. In slow-roll inflation, the rolling inflaton background breaks the de Sitter boosts, and the interactions it induces are not $\mathrm{SO}(d+2)$ invariant. It would be interesting to extend our framework to such backgrounds and compute phenomenologically relevant observables.
\end{itemize}

\section*{Acknowledgments}
We would like to thank Dio Anninos, Shounak De, Frederik Denef, Aidan Herderschee, Atsushi Higuchi, Hayden Lee, Juan M. Maldacena, Enrico Pajer, Jo\~ao Penedones, Guilherme  Pimentel,  Charlotte Sleight,  Massimo Taronna, and Zihan Zhou for useful discussions. LLMs, OpenAI GPT Astra/Sol and Anthropic Claude Fable/Opus, were used for editing and for independent checks of equations. ZS is supported by the U.S. Department of Energy grant DESC0009988 and the Sivian Fund.

\appendix

\section{Derivative interactions and massless scalars}\label{derivative}
It is straightforward to incorporate interaction vertices with derivatives in the sphere  formalism. As an example, we add two derivatives to the $n$-point contact diagram:
\begin{equation}
    \CI_{\partial}=\int_{S^{d+1}} \nabla_a G_{\Delta_1}(\hat X_1, \hat X_0)\nabla^a G_{\Delta_2}(\hat X_2, \hat X_0)\prod_{k=3}^nG_{\Delta_k}(\hat X_k, \hat X_0)~,
\end{equation}
where $\nabla_a$ is the covariant derivative in some local coordinates on the sphere at point $\hat X_0$.  In the embedding space representation of $G$, the derivatives lead to the following term:
\begin{equation}\label{add1}
    (2\mu_1^2 X_0\cdot\nabla_a X_0-2\lambda_1\mu_1 \hat X_1\cdot \nabla_a X_0)(2\mu_2^2 X_0\cdot\nabla^a X_0-2\lambda_2\mu_2\hat X_2\cdot \nabla^a X_0)~,
\end{equation}
where ``$\cdot$'' denotes contraction of embedding space indices. \eqref{add1} can be drastically simplified by noting that $\nabla_a X_0^I \nabla^a X_0^J = X_0^2 \delta^{IJ} - X_0^I X_0^J$. This embedding space tensor annihilates $X_0$, implying that $ X_0\cdot\nabla_a X_0$ in \eqref{add1} does not contribute. The remaining term becomes $4\lambda_1\mu_1\lambda_2\mu_2 (X_0^2 \hat X_1\cdot \hat X_2 - X_0\cdot \hat X_1 X_0\cdot \hat X_2)$. The contribution of $4\lambda_1\mu_1\lambda_2\mu_2 X_0^2 \hat X_1\cdot \hat X_2$ is relatively easy. Following the same procedure as in the main text, we obtain its  contribution to the boundary correlator
\begin{equation}
    \CI^{\dS}_{\partial, 1}=\frac{-2\vec y_{12}^2\prod_i \Gamma(\frac{d}{2}-\Delta_i)}{2^n (2\pi^{\frac{\bD}{2}})^{n-1}\Gamma\left(\frac{\bD-\Delta_{\rm tot}}{2}-1\right)}D_{\Delta_1+1,\Delta_2+1,\Delta_3,\cdots,\Delta_n} (\vec y)\,.
\end{equation}

For the contribution of the other term, we  notice that inserting $\lambda_1\partial_{\lambda_1}$ in the integral representation gives $-2\lambda_1^2 + 2\lambda_1\mu_1 \hat X_1\cdot X_0$. So the combination $4\lambda_1\mu_1\lambda_2\mu_2 X_0\cdot \hat X_1 X_0\cdot \hat X_2$ is equivalent to $(\lambda_1\partial_{\lambda_1}+2\lambda_1^2) (\lambda_2\partial_{\lambda_2}+2\lambda_2^2)$, where the derivative $\lambda_i\partial_{\lambda_i}$ can be replaced by $-\bar\Delta_i$ as a result of integration by parts. Altogether, we just need to insert an additional factor of $(2\lambda_1^2-\bar\Delta_1)(2\lambda_2^2-\bar\Delta_2)$ in the non-derivative case, which leads to 
\begin{equation}
    \CI^{\rm dS}_{\partial, 2} =\frac{\Delta_1\Delta_2\prod_i \Gamma(\frac{d}{2}-\Delta_i)}{2^n (2\pi^{\frac{\bD}{2}})^{n-1}\Gamma\left(\frac{\bD-\Delta_{\rm tot}}{2}\right)}D_{\Delta_1,\Delta_2,\cdots,\Delta_n}(\vec y)\,.
\end{equation}
The final result for this contact diagram with derivatives is thus $\CI^{\rm dS}_{\partial} =\CI^{\rm dS}_{\partial, 1} -\CI^{\rm dS}_{\partial, 2} $.

Both $\CI^{\rm dS}_{\partial, 1} $ and $\CI^{\rm dS}_{\partial, 2} $ have a well-defined massless limit for external legs attached to $\hat X_1$ and $\hat X_2$, i.e., $\Delta_1=\Delta_2=0$. For $\CI^{\rm dS}_{\partial, 1} $, this is a trivial claim because the $D$ function itself is well defined in this limit. For $\CI^{\rm dS}_{\partial, 2} $, we encounter an integral of the form 
\begin{equation}
    F = \Delta_1\Delta_2 \int_\mu \mu_1^{\Delta_1}\mu_2^{\Delta_2} e^{-\mu_1\mu_2 A - (\mu_1 B+\mu_2 C)}
\end{equation}
and we need the leading behavior in the $\Delta_1, \Delta_2\to 0$ limit. Performing the $\mu_1$ integral yields $\Gamma(\Delta_1+1)/(B+\mu_2 A)^{\Delta_1}$, where we have included the prefactor $\Delta_1$. This function approaches 1 in the $\Delta_1\to 0$ limit. The remaining integral of $\mu_2$ leads to $\Gamma(\Delta_2+1)/C^{\Delta_2}$, which approaches 1 in the $\Delta_2\to 0$ limit. 
It means that in the massless limit, we can get rid of the Schwinger parameters $\mu_1$ and $\mu_2$ in $\CI^{\rm dS}_{\partial, 2} $.

\section{Mellin amplitude of the dS exchange diagram }\label{mellin}
In this appendix, we derive a MB representation for \eqref{ss}.
First, it is known that the $\mu_i$ integrals in \eqref{ss} admit a MB representation as follows \cite{Symanzik:1972wj, Penedones:2010ue}
\begin{equation}
    \int_\mu \prod_i\mu_i^{\Delta_i} e^{-\Phi_{\rm ex} } = \frac{1}{2}\int \frac{d\delta_{12}d\delta_{14}}{(2\pi i)^2}\frac{\prod_{1\le i<j\le 4}\Gamma(\delta_{ij}) \vec y_{ij}^{-2\delta_{ij}}}{(1+\sigma^2)^{\delta_{12}}(1+\rho^2)^{\delta_{34}}(\rho\sigma)^{\delta_{13}+\delta_{14}+\delta_{23}+\delta_{24}}}~,
\end{equation}
where $\delta_{ij}=\delta_{ji}$ satisfy $\sum_{j\not=i}\delta_{ij}=\Delta_i$. There are 4 constraints for 6 variables. We choose the independent variables to be $\delta_{12}$ and $\delta_{14}$. Some useful consequences of the constraints are 
\begin{equation}
    \delta_{13}+\delta_{14}+\delta_{23}+\delta_{24} = \Delta_1+\Delta_2-2\delta_{12}=\Delta_3+\Delta_4-2\delta_{34}~.
\end{equation}
Therefore the $\rho$ and $\sigma$ integrals only depend on $\delta_{12}$. It is also conventional to define the variables $s_{12}$ and $s_{14}$ by
\begin{equation}
    \delta_{12} = \frac{\Delta_1+\Delta_2-s_{12}}{2}, \quad \delta_{14} = \frac{\Delta_1+\Delta_4-s_{14}}{2}~.
\end{equation}
After making these changes of variables and the substitution $(\rho,\sigma)\to (\sqrt{\rho},\sqrt{\sigma})$, the integrals in \eqref{ss} become
\begin{equation}
    \begin{aligned}
    &\int_{\rho, \sigma}\frac{\rho^{\bar\Delta}\sigma^\Delta }{(1+\rho^2+\sigma^2)^{\frac{\bD-\Delta_{\rm tot}}{2}}}\int_\mu \mu_i^{\Delta_i} e^{-\Phi_{\rm ex} }\\
    &= \frac{1}{8}\int \frac{d\delta_{12}d\delta_{14}}{(2\pi i)^2}\prod_{i<j}\frac{\Gamma(\delta_{ij})}{\vec y^{2\delta_{ij}}_{ij}}\int_{\rho, \sigma}\frac{\rho^{\frac{\bar\Delta-s_{12}}{2}}\sigma^{\frac{\Delta-s_{12}}{2}} }{(1+\rho+\sigma)^{\frac{\bD-\Delta_{\rm tot}}{2}}(1+\sigma)^{\frac{\Delta_1+\Delta_2-s_{12}}{2}}(1+\rho)^{\frac{\Delta_3+\Delta_4-s_{12}}{2}}}\,.
\end{aligned}
\end{equation}
Then we introduce an additional MB integral to decouple $\rho$ and $\sigma$
\begin{equation}
    \frac{\Gamma(\frac{\bD-\Delta_{\rm tot}}{2})}{(1+\rho+\sigma)^{\frac{\bD-\Delta_{\rm tot}}{2}}} = \int\frac{dz}{2\pi i}\Gamma(-z)\Gamma\left(\frac{\bD-\Delta_{\rm tot}}{2}+z\right) \frac{\sigma^z}{(1+\rho)^{\frac{\bD-\Delta_{\rm tot}}{2}+z}}\,,
\end{equation}
which allows us to integrate out $\rho$ and $\sigma$ separately
\begin{equation}\label{sss}
    \begin{aligned}
        &\int_{\rho, \sigma}\frac{\rho^{\frac{\bar\Delta-s_{12}}{2}}\sigma^{\frac{\Delta-s_{12}}{2}} }{(1+\rho+\sigma)^{\frac{\bD-\Delta_{\rm tot}}{2}}(1+\sigma)^{\frac{\Delta_1+\Delta_2-s_{12}}{2}}(1+\rho)^{\frac{\Delta_3+\Delta_4-s_{12}}{2}}}\\
        &=\frac{\Gamma(\frac{\bar\Delta-s_{12}}{2})}{\Gamma(\frac{\Delta_1+\Delta_2-s_{12}}{2})}\int\frac{dz}{2\pi i}\frac{\Gamma(-z)\Gamma(\frac{\Delta-s_{12}}{2}+z)\Gamma\left(\frac{\bD-\Delta_{\rm tot}}{2}+z\right)\Gamma(\frac{\Delta-\Delta_1-\Delta_2}{2}+z+1)\Gamma(\frac{\Delta_1+\Delta_2-\Delta}{2}-z)}{\Gamma(\frac{\bD-\Delta_{\rm tot}}{2})\Gamma(\frac{\bD-\Delta_1-\Delta_2-s_{12}}{2}+z)}\,. 
    \end{aligned}
\end{equation}
This MB integral defines a function of $s_{12}$ with poles at $s_{12}\in\{\Delta, \bar\Delta\}+2\mathbb Z_{\ge 0}$. The poles at $ \bar\Delta+2\mathbb Z_{\ge 0}$ can be easily seen from the prefactor $\Gamma(\frac{\bar\Delta-s_{12}}{2})$, and the poles at $ \Delta+2\mathbb Z_{\ge 0}$ arise from the collision of poles (in the $z$-plane) of $\Gamma(-z)\Gamma(\frac{\Delta-s_{12}}{2}+z)$. Because of the $\Delta\leftrightarrow\bar\Delta$ symmetry of \eqref{sss},
it suffices to  compute the residue at $s_{12}= \bar\Delta+2n$, where $\Gamma(\frac{\bar\Delta-s_{12}}{2})$ has a simple pole with  residue  $2(-1)^{n+1}/n!$. The $z$ integral can be evaluated analytically at these values of $s_{12}$ using the first Barnes lemma, leading to 
\begin{equation}
   {\rm Res}_{s_{12}=\bar\Delta+2n}\eqref{sss}= 2(-)^{n+1}\frac{\Gamma\left(\frac{\Delta-\bar\Delta}{2}-n\right)\Gamma\left(\frac{\bar\Delta-\Delta_3-\Delta_4}{2}+1+n\right)\left(\frac{\bar\Delta-\Delta_1-\Delta_2}{2}+1\right)_n}{n!\Gamma\left(\frac{\Delta-\Delta_3-\Delta_4}{2}+1\right)}\,.
\end{equation}
With the knowledge about the analytical structure, we can reconstruct \eqref{sss} by summing up the poles. Altogether, the full MB representation of \eqref{ss} is 
\begin{equation}\label{Mder}
  \langle\CO_{\Delta_1}(\vec y_1)\cdots \CO_{\Delta_4}(\vec y_4)\rangle^{\SphereAffil{}}_{\rm ex} = \int \frac{d\delta_{12}d\delta_{14}}{(2\pi i)^2} \CM^{\rm dS}_{\rm ex}(s_{12})\prod_{i<j}\frac{\Gamma(\delta_{ij})}{\vec y^{2\delta_{ij}}_{ij}}\,,
\end{equation}
where the Mellin amplitude $\CM^{\rm dS}_{\rm ex}$ reads
\begin{equation}\label{Mex}
    \CM^{\rm dS}_{\rm ex} = \frac{\pi \prod_i \Gamma(\frac{d}{2}-\Delta_i)}{2^5(2\pi^{\frac{\bD}{2}})^3}\sum_{n\ge 0}-\frac{c_{\bar\Delta}(\frac{\bar\Delta-\Delta_1-\Delta_2}{2}+1)_n(\frac{\bar\Delta-\Delta_3-\Delta_4}{2}+1)_n}{\Gamma(\frac{\Delta\!-\!\Delta_1\!-\!\Delta_2}{2}\!+\!1)\Gamma(\frac{\Delta\!-\!\Delta_3\!-\!\Delta_4}{2}\!+\!1)\Gamma(\frac{\bar\Delta\!-\!\Delta}{2}\!+\!1\!+\!n)n!}\frac{1}{s_{12}\!-\!\bar\Delta\!-\!2n}+(\bar\Delta\leftrightarrow\Delta)\,.
\end{equation}

On the other hand, the $s$-channel exchange Witten diagram in EAdS depends on the boundary condition of the internal line. With the choice of standard quantization for the internal line, the Mellin amplitude of the Witten diagram is \cite{Penedones:2010ue} 
\begin{equation}\label{MAdS}
    \CM^{\rm AdS}_{\rm ex, \Delta} = \frac{\Gamma(\frac{\Delta_1+\Delta_2-\bar\Delta}{2})\Gamma(\frac{\Delta_3+\Delta_4-\bar\Delta}{2})}{8(2\pi^{\frac{d}{2}})^3\prod_i\Gamma(\Delta_i-\frac{d}{2}+1)}\sum_{n\ge 0} -\frac{(\frac{\Delta-\Delta_1-\Delta_2}{2}+1)_n(\frac{\Delta-\Delta_3-\Delta_4}{2}+1)_n}{n!\Gamma(\frac{\Delta-\bar\Delta}{2}\!+\!1\!+\!n)} \frac{1}{s_{12}-\Delta-2n}\,,
\end{equation}
and it can be analytically continued to more general values of $\Delta$ in the complex plane. Comparing~\eqref{Mex} with~\eqref{MAdS}, we find that the dS Mellin amplitude $\mathcal{M}^{\rm dS}_{\rm ex}$ can be expressed as a linear combination of the corresponding AdS Mellin amplitudes, with the internal line subject to either 
$\Delta$ or $\bar\Delta$ boundary condition:
\begin{equation}
    \CM^{\rm dS}_{\rm ex} = 4\left(\prod_{i}c_{\Delta_i}\right)\left(c_{\Delta}\sin\left(\frac{\Delta_1\!+\!\Delta_2\!-\!\bar\Delta}{2}\pi\right)\sin\left(\frac{\Delta_3\!+\!\Delta_4\!-\!\bar\Delta}{2}\pi\right) \CM^{\rm AdS}_{\rm ex, \Delta}+(\Delta\leftrightarrow\bar\Delta)\right)\,,
\end{equation}
where $c_\Delta$ is defined by \eqref{contactpos}.

\iflatexml
\providecommand{\href}[2]{#2}\begingroup\raggedright\endgroup

\else
\bibliographystyle{JHEP}
\bibliography{refs.bib}
\fi

\end{document}